\documentclass[final,authoryear,5p]{elsarticle}

\usepackage{graphicx}

\usepackage{amssymb}
\usepackage{amsmath}

\usepackage[ps2pdf,%
a4paper=true,%
breaklinks=true,%
colorlinks=true,%
pdfauthor={Boschini et al.},%
pdftitle={Propagation of Cosmic Rays in Heliosphere: the HelMod Model}%
]{hyperref}

\journal{Advances in Space Research}

\biboptions{authoryear}

\newcommand{\degree}{\ensuremath{^\circ }}
\newcommand{\galprop}{GALPROP}
\newcommand{\helmod}{\textsc{HelMod}}

\newcommand{\apj}{Astrophys. J.}
\newcommand{\apjl}{Astrophys. J. Letter}
\newcommand{\nat}{Nature}
\newcommand{\apss}{Astrophysics and Space Science}
\newcommand{\prd}{Phys. Rev. D}
\newcommand{\solphys}{Sol. Phys.}
\newcommand{\jgr}{J. Geoph. Res.}

\begin{document}

\begin{frontmatter}


\title{Propagation of Cosmic Rays in Heliosphere: the \helmod{} Model}

\author[label1,label2]{M.~J.~Boschini}
\author[label1]{S.~{Della~Torre}}
\author[label1,label3]{M.~Gervasi}
\author[label1]{G.~{La~Vacca}}
\author[label1]{P.~G.~Rancoita\corref{cor}}
\cortext[cor]{Corresponding author}
\ead{piergiorgio.rancoita@mib.infn.it}
\address[label1]{INFN sez. Milano-Bicocca, Piazza della Scienza, 3 - 20126 Milano (Italy)}
\address[label2]{also CINECA, Segrate, Milano, Italy}
\address[label3]{also Physics Department, University of Milano-Bicocca, Piazza della Scienza, 3 - 20126 Milano (Italy)}

\begin{abstract}
The heliospheric modulation model \helmod{} is a two dimensional treatment dealing with the helio-colatitude and radial
distance from Sun and is employed to solve the transport-equation for the GCR propagation through the heliosphere down to Earth.
This work presents the current version 3 of the \helmod{} model and reviews how main processes involved in GCR propagation were implemented.~The treatment includes the so-called particle drift effects --~e.g., those resulting, for instance, from the extension of the neutral current sheet inside the heliosphere and from the curvature and gradient of the IMF --, which affect the
transport of particles entering the solar cavity as a function of their charge sign.~The \helmod{} model is capable to provide modulated spectra which well agree within the experimental errors with those measured by AMS-01, BESS, PAMELA and AMS-02 during the solar cycles 23 and 24.~Furthermore, the counting rate measured by Ulysses at $\pm 80\degree$ of solar latitude and 1 to 5\,AU was also found in agreement with that expected by \helmod{} code version 3.
\end{abstract}


\end{frontmatter}

\parindent=0.5 cm

\section{Introduction}\label{Introduction}
\label{Intr}
Modulated omni-directional intensities of galactic cosmic rays (GCRs)
were observed during different phases of solar activity using both balloon flights~\citep[for instance, see][]{BoezioetAl1999,Mennetal2000,HainoSanuki2004,bess_prot,AbeetAl2008,BESS2007_Abe_2016} and space-borne missions~\citep[e.g., see][]{AMS_protons,AMS_leptons,AMS_cosmic,AMS_helium,AMS01_prot,AMS_positron,AdrianietAl2009a,AdrianietAl2009b,AdrianietAl2010,PAMELA_Prot_He_2011,PamelaProt2013,2015ApJ...810..142A,2016PhRvL.116x1105A,AMS02_2014_PhysRevLett2,AMS02_2015_PhysRevLett1,AMS02_2015_PhysRevLett2,AMS02_2016_PhysRevLett1}, in particular during the latest solar cycles.~The increased performance of on-board spectrometers was and is currently enabling to enhance the accuracy of the observed spectra.~Thus, it was opening the way to
a better understanding of processes related to the transport of GCRs
through the Heliosphere -- the so-called \textit{modulation effect} -- and, ultimately, to the capability a) to unveil
local interstellar spectra (LIS)
of GCR species{~\citep[e.g., see][and also references therein]{2014ApJ...794..166B,2016Ap&SS.361...48B,ECRS2016b,LIS_ApJ2017}}, b)
to investigate their generation, acceleration and diffusion process
within the Milky Way \citep[e.g., see][]{Boella1998,StrongMoskalenko2007,EvoliGAggero2008,PutzeMaurin2009}, and, in turn, c)
to possibly untangle features due to new physics
-- i.e.,~dark matter \citep[e.g., see][and references therein]{BottinoDonato1998,CirelliCline2010,IbarraTran2010,Salati2011,Weniger2011} --  or additional astrophysical sources so far not taken into account \citep[e.g., see][and references therein]{Chang2008, AbdoetAl2009,AdrianietAl2009a,Cernuda2011,MertschSarkar2011,RozzaJHEA_2015,ICRC2015_Rozza}.
\par
Among space missions currently observing GCRs,
AMS-02 -- on-board of the International Space Station since May 2011 -- is continuously providing data with unprecedented measurement accuracy.~In fact, this spectrometer allowed one to determine the most accurate differential intensities of protons~\citep{AMS02_2015_PhysRevLett1}, helium nuclei~\citep{AMS02_2015_PhysRevLett2}, antiprotons~\citep{AMS02_2016_PhysRevLett1}, electrons and positrons~\citep{AMS02_2014_PhysRevLett2}.~The high precision of these experimental data together with those from Ulysses spacecraft~\citep[e.g., see][]{Simpson1992,Simpson1996b,Heber1996,FerrandoEtAl1996,deSimone2011,GieselerHeber2016} constitute a challenge for any modulation model of the
\textit{inner part of heliosphere}. 
Actually such a treatment has to reproduce the observed GCR spectra transported -- during different solar activity phases -- down to Earth and, also, outside the ecliptic plane at distances from Sun ranging from about 1 to 5\,AU.~In fact, the observations made using Ulysses spacecraft
allowed one to determine both the latitudinal and radial dependence of GCR intensity. Furthermore, the data taken during Ulysses fast latitudinal scan exhibited a latitudinal dependence on i) the
charge sign of the GCR species (i.e., protons and electrons, which are the dominant positively and negatively charged component, respectively), ii) solar activity and iii) polarity of the interplanetary magnetic field (IMF).~It is worth to remark that these data may, in addition, allow a better understanding of space radiation environment close to Earth, thus extending our capability of predicting radiation hazards for astronauts
and device damages in space missions~{\citep[e.g., see][and Chapters 7 and 8 of~\cite{rancoita2015}]{leroyRancoita2007,ICRCDamage2015}}.
\par
In this work we present the version 3 of 2-D heliospheric modulation (\helmod{}) model --~i.e., a two dimensional treatment dealing with the helio-colatitude and radial
distance from Sun~\citep{GervasiEtAl1999,Bobik2011ApJ,DellaTorre2013AdvAstro} -- currently employed to solve the transport-equation for the GCR propagation through the heliosphere down to Earth.
{The relevant GCR propagation processes are described in Sects.~\ref{Model} and \ref{SEC::Parameters} in order to illustrate how they are implemented in his latest version of \helmod{} model. Furthermore, details on Monte Carlo technique used to solve the stochastic integration are treated in Sects. \ref{Code}, \ref{Results}, and allow one a better understanding  on \helmod{} capabilities to deal with solar modulation, within the inner part of the heliosphere.}
At present, the model only treats GCRs with energies $\gtrsim 0.5 $\,GeV/nucleon, thus modulation effects occurring in the outer heliosphere -- i.e., beyond the termination shock (TS)~\citep[see, e.g.,][]{langner2003,Langner2004,Bobiketal2008,Potgieter_2008,2009ApJ...701..642F,2013ApJ...764...85L,2013ApJ...778..122S} -- are not accounted for.
~It has to be pointed out that the \helmod{} model is capable of describing the current large set of observation data, which were collected during solar cycles 23 and 24 with the occurrence of two solar minimum.~For this purpose, the model includes the so-called \textit{particle drift effects} --~e.g., those resulting, for instance, from the extension of the neutral current sheet inside the heliosphere and from the curvature and gradient of the IMF --, which affect the
transport of particles entering the solar cavity as a function of their charge sign.~These effects are particularly relevant when IMF exhibits a well-defined large-scale structure or this latter is still relevant.~In fact, at the solar minimum and when the solar activity is not too far from such a condition, GCR modulated intensities exhibit a dependence on charge sign~\citep[e.g., see][]{GarciaMeyer1986,ClemEtAl1996,ClemEvenson2000,BoellaEtAl2001}.
In fact, the IMF polarity reversal causes charge sign
dependent modulation effects, for instance, those
observed in particle over anti-particle intensities
ratio at rigidities lower than about 10--20 GV~\citep[e.g.,][]{2016PhRvL.116x1105A}. These
effects are treated in the Parker transport equation
through the terms including the drift velocity.
The analysis on Ulysses out-of-ecliptic observations~\citep[e.g., see][]{Simpson1996a,Simpson1996b,Heber1996,Heber1998,Heber2008,FerrandoEtAl1996,deSimone2011,GieselerHeber2016} provided, so far, a unique point of view highlighting the presence of latitudinal gradients in the spatial distribution of GCRs, during period of low solar activity,~i.e., when the combination of particle charge ($q$) and solar magnetic polarity ($A$) is positive ($qA>0$); while a more uniform distribution of GCRs in the inner part of the heliosphere occurs for $qA<0$.
\par
As discussed in~\citet{AstraArticle2011,Bobik2011ApJ}, the model exhibits a smooth time dependence introduced by the parameters -- related to solar activity and adopted within the model itself, as described in Sect.~\ref{Sec::K0Fit} --,
which are averaged over time durations corresponding to Carrington rotations, i.e., a) solar wind speed ($V_{\rm sw}${, see Sect.~\ref{SEC::kpar}}), b) tilt angle ($\alpha_{\rm t}${, see Sect.~\ref{SEC::Parameters}}) of the neutral current sheet, and c) diffusion parameter ($K_{0}${, see Sect.~\ref{Sec::K0Fit}}).~Furthermore, it has to be remarked that the solar wind usually takes one year or even more to reach the border of heliosphere.~The above parameters -- usually determined at 1 AU -- are transferred to describe the properties of any distant heliospheric sector, according to the time required by the solar wind to reach such a region from Sun~{(see discussion in Sect.~\ref{SEC::Parameters})}.~
\par
In the present article, the LIS fluxes are those derived in~\citet{ECRS2016b,LIS_ApJ2017} by means of GALPROP v55.
In that article leptons were not yet treated, i.e., no electron LIS is available yet using the new GALPROP version.
Thus, the discussion on the modulated spectra obtained using \helmod{} model is mostly restricted to comparisons with experimental data regarding protons and helium nuclei. As discussed in~\citet{ECRS2016b,LIS_ApJ2017}, the LIS, presented in Sect.~\ref{SEC::Parameters}, accommodate both the low energy interstellar CR spectra measured by Voyager 1 and the high energy observations by BESS, Pamela, AMS-01, and AMS-02 over solar cycles 23--24.
\par
Finally, we have to remark that \citet{2013ApJ...779..158E} exploited an \textit{ab-initio} approach for a three dimensional steady state GCR modulation model, in which the effects of turbulence on both the diffusion and drift of these cosmic-rays are treated in a self-consistent description{; \citet{2013ApJ...765L..18S} uses a hybrid modeling approach incorporating the  plasma flow from a magnetohydrodynamics model with the particle transport; and, finally, \citet[and reference therein]{2016SoPh..291.2181V} computed spatial gradients and absolute flux variations for GCR protons in the heliosphere for solar minimum}.~Although these models provide encouraging results, they still depends on parameters whose time evolution is not yet measurable or fully understood.~So far, the found agreement among the modulated spectra from \helmod{} code and experimental data collected over a long period~\citep[e.g., see][]{Bobik2011ApJ,DellaTorre2013AdvAstro,ECRS2016b,LIS_ApJ2017} motivated the choice, in \helmod{}, to reduce
the
complexity of diffusion process using a unique time dependent variable, as described in Sect.~\ref{SEC::kpar}.


%
\section{Heliospheric Propagation of Cosmic Rays}
\label{Model}
%
\subsection{Parker Equation}\label{SEC:parkerEq}
The cosmic rays propagation trough the heliosphere was treated by~\citet{parker1965}, who demonstrated that - in the framework of statistical physics -
the random walk of cosmic ray particles is a Markoff process, describable by a Fokker--Planck equation (FPE).~
In his original formulation, Parker's transport-equation was expressed in terms of \textit{particle density} for unit space and energy,~i.e.,~$U(\vec{x},T)$~{\citep[e.g., see][and also Sections 8.2--8.2.5 of~\citealp{rancoita2015}]{Jokipii1970,Fisk1971,Bobik2011ApJ}}:
\begin{equation}\label{eq_parker}
 \frac{\partial U}{\partial t} = - \nabla \cdot (U\vec{V})
                                 + \nabla \cdot \left[  K^S \cdot \nabla U\right]
                                 + \frac{(\nabla\cdot\vec V_{sw})}{3}\frac{\partial}{\partial T}\left(\alpha_{rel} T U \right),
\end{equation}
with $\vec{x}$ the 3D-spatial position in Cartesian coordinates, \[\vec{V} = \vec{V}_{sw} + \vec{v}_{drift},\] $\vec{V}_{sw}$ the
solar wind (SW) velocity,
\begin{equation}\label{drift_antisym_ten}
     \vec{v}_{drift}  =  \nabla \cdot  K^A
\end{equation}
the drift velocity~\cite[e.g., see][and references therein]{Jokipii77,JokipiLev1977, Bobik2011ApJ}, $ K^A$ and $ K^S$ the antisymmetric and symmetric part of the diffusion tensor, respectively.~In Eq.~(\ref{eq_parker})
$T$ is particle kinetic energy, $T_0$ is particle rest energy  and, finally, \[\alpha_{rel} = \frac{T + 2T_{0}}{T+T_{0}}.\]
We should note that some authors prefer to describe the modulation of GCRs re-expressing Eq.~(\ref{eq_parker}) in terms of the so-called
\textit{omni-directional distribution function} $f(\vec{x},p)$, where $p$ is particle momentum~\citep[e.g., see][and Section 8.2.4 of~\citealp{rancoita2015}]{JokipiiKopriva1979,Yamada1998,PeiBurger2010}:
\begin{equation}\label{eq_parker_p}
 \frac{\partial f}{\partial t} = - \nabla \cdot (f\vec{V})
                                 + \nabla \cdot \left[  K^S \cdot \nabla f\right]
                                 + \frac{(\nabla\cdot\vec V_{sw})}{3  p^2}\frac{\partial}{ \partial p}\left(p^3 f \right).
\end{equation}
Although Eqs.~\eqref{eq_parker} and~\eqref{eq_parker_p} describe the same process~\citep[see, e.g.,][]{2002cra..book.....S}, they are commonly used as alternative formulation; therefore the SDE integration technique (described in Sect.~\ref{Code}) used in this work  should be performed differently as described in~\citet{DellaTorre2016_OneD}.
\par
Parker's transport-equation describes
i) the \textit{diffusion} of GCRs by magnetic irregularities,
ii) the so-called \textit{adiabatic-energy changes} associated with expansions and compressions of cosmic radiation,
iii) an \textit{effective convection} resulting from the \textit{solar wind} (SW, with velocity $\vec{V}_{{\rm sw}}$) \textit{convection} effect
and iv) the drift effects related to the \textit{drift velocity} ($\vec{v}_{drift}$).
In turn, the drift velocity is determined by the antisymmetric part of the diffusion tensor [see Eq.~(\ref{drift_antisym_ten})]
which accounts for gradient, curvature and current sheet drifts of particles in the IMF,
i.e., it depends on the charge sign of particles.
\par
{ As discussed in Sect.~\ref{SEC::kpar}, the low scale irregularities
are the main responsible of particle diffusion in the heliosphere.~Nevertheless, large scale structures due to SW expansion contribute
to convective and magnetic drift motion in Eq.~\eqref{eq_parker}.~Although present in the original formulation of Parker's equation, in early modulation models the drift velocity was neglected.~In recent years, it was found that models accounting for charge dependent effect by means of drift velocity are, indeed, capable to describe the observations~\citep[e.g., see][]{DellaTorre2012,ICRC13_DellaTorre,2013PhRvL.110h1101M}.

In \helmod{} model, we use the drift treatment originally developed by~\citet{Potgieter85}
and refined using Parker's magnetic field with polar correction described in Sect.~\ref{SEC::HMF}. The full description and derivation can be found in~\citet{DellaTorre2013AdvAstro}.~Since during high activity periods the heliospheric magnetic field is far from
being considered regular, we introduced a correction factor suppressing any drift velocity at solar maximum.~For sake of completeness, we have to note that the presence of turbulence
in the interplanetary medium should reduce the global effect of CR drift in the heliosphere
(e.g., see discussion in~\citealt{Minnie2007}) and this is usually
taken into account introducing a \textit{drift suppression factor}~\citep[e.g., see][]{Strauss2011} that is effective at rigidities below 1 GV~\citep[for a general discussion about the energy range of drift effects in the solar modulation see, e.g.,][]{2016AdSpR..58..453N}.
\par
As remarked by \citet{Jokipii77}, since $\nabla \cdot \vec{v}_{drift} = 0$, one finds that drift velocity is added to that of solar wind
resulting \textit{effective convection speed} but do not contribute to the adiabatic-energy losses [third right-hand term of Eqs.~(\ref{eq_parker},~\ref{eq_parker_p})]. Nevertheless, his contribution causes an interplay with the diffusion process resulting in different modulated spectral slope for period of same solar activity but opposite magnetic field polarity~\citep[see, e.g.,][]{2013LRSP...10....3P}.
\par
In \helmod{} model version 3, the description of the solar wind is that presented in~\citet{Bobik2011ApJ}, i.e., it is constant  (from Earth up to the TS) and its value is that measured at Earth.
During periods of low solar activity, the solar wind speed increases by
almost a factor two from the ecliptic plane to the poles, thus subdividing the heliosphere into two regions with slow and fast
solar wind~\citep{McComas2000}; no latitudinal dependence is considered during high activity periods~\citep{Bobik2011ApJ}.

We should note that second-order Fermi acceleration mechanisms are usually neglected in general treatment
of the propagation of GCRs towards the inner part of heliosphere, while additional energy-loss processes --
beside that adiabatic included in Eq.~(\ref{eq_parker}) -- (like e.g. inverse-Compton) are negligible~\citep[e.g., see][]{ICATPP10_energyloss}.
\par
Finally, the differential intensity of GCRs, $J$, is related to the particle density ($U$) and to the omni-directional distribution function ($f$) [see Eqs.~(\ref{eq_parker},~\ref{eq_parker_p})] as:
\begin{equation}\label{eq::DiffIntensity}
 J = p^{2}f= \frac{\beta\, c \,U}{4 \pi},
\end{equation}
where $\beta$ is the particle velocity expressed in units of the speed of light, $c$.
\subsection{The Diffusion Tensor}
\label{SEC::kpar}
When~\citet{parker1965} introduced his transport equation,
he underlined how the diffusion process experienced by cosmic rays is mainly resulting from the magnetic irregularities of the IMF.~Such magnetic-field irregularities, he remarked in that work, are transported rigidly in the SW and appear with dimensions comparable with the gyro-radius of about 0.1--10\,GeV proton moving in an IMF of a few nT.~The scattering of energetic particles on them causes a random walk, which is accounted for by a diffusion tensor having a parallel component ($K_{||}$) -- with respect to the direction of large scale IMF -- larger than that perpendicular [$K_{\perp,i}$, where the subscript $i$ refer to radial ($r$) or latitudinal ($\theta$) direction components].~For rigidity greater than 1\,GV, the \helmod{} code version 3 implements a functional form that is linearly dependent to particle rigidity ($P$)
and linearly proportional to solar distance ($r$)~\citep[e.g., see][]{LIS_ApJ2017}:
\begin{equation}\label{EQ::KparActual}
 K_{||}=\frac{\beta}{3} K_0\left[ \frac{P}{1\textrm{GV}}+g_{low}\right] \left(1+\frac{r}{\textrm{1\,AU}}\right),
\end{equation}
where $K_0$ is the diffusion parameter --  described in Sect.~\ref{Sec::K0Fit} --
{which dependence on time reflect the variability of interplanetary medium properties for the different} phases of solar activity,
\[P=\frac{p\,c}{|Z|e}\] is the \textit{particle rigidity} expressed in GV with $p$ the particle momentum, $r$ is the
heliocentric distance from Sun in AU and, finally, $g_{low}$ is a parameter {(see a further discussion in this Section)}, which depends on the level of solar activity and allows the description of the flattening with rigidity below few GV.~In the present model, the spatial dependence is proportional to the distance ($r$) from Sun; it is consistent with that used in~\citet{DellaTorre2013AdvAstro}
for dealing with CR latitudinal gradients and no further latitudinal dependence appears to be needed~\citep[see also discussions in][]{JokipiiKota89,McDonald1997,Strauss2011}.~
\par
It is important to remark how from the observation of fluctuations in the average magnetic field,~\citet{jokipii1966} and \citet{jokipii1971} put the basis for the first successful description of diffusion tensor with the
so-called \textit{quasi-linear theory }(QLT) under the approximation of a weak turbulence.
However, a complete parametrization  
 for the components of diffusion tensor, from low to high rigidities, are still an open question.~Nowadays, it is commonly accepted that~\citep[see, e.g.,][and reference therein]{shalchi2009,Engelbrecht2015}:
i) at higher rigidities the diffusion coefficient should have a quasi-linear dependence~\citep[e.g., see][]{Gloeckler1966,FFM1968,jokipii1966,jokipii1971,perko1987,potgieter1994,Strauss2011}, and ii) the diffusion coefficient should be more ``flat'' at lower rigidity~\citep[e.g., see][]{Palmer1982,Bieberetal1994}.~In previous models of CR propagation in the heliosphere, at intermediate energies, to account for the above discussed conditions the parallel diffusion coefficient ($K_{||}$) was expressed with a sharp transition at $\sim 1$\,GV between the two above mentioned \textit{regimes}~\citep[e.g., see][]{perko1987,alanko2007,Strauss2011,Bobik2011ApJ}.~However, with increasing the experimental accuracy of collected data such a simple approach needs to be revised allowing a smoother transition between the two \textit{regimes}.~In fact, the present functional form of such a transition  [Eq.~\eqref{EQ::KparActual}] is consistent with those presented in~\citet{BurgerHattingh1998} for the same rigidity interval.
\par
In the current model the perpendicular diffusion coefficient is taken to be proportional to $K_{||}$ following the ratio
\[\frac{K_{\perp,i}}{K_{||}}=\rho_i,\] for both $r$ and $\theta$ $i$-coordinates\citep[e.g., see ][and references therein]{potgieter2000, BurgerHattingh1998,LIS_ApJ2017}.
~\citet{Palmer1982} suggested that \[0.022 < \rho_i< 0.083\] at Earth.~The above description for $K_{\perp,i}$ is consistent at high rigidity with those from QLTs.
Although this description was improved using more complicated approach~\citep[see, e.g.,][]{2013ApJ...779..158E}, it remains one of the fundamental observational reference for transport theories~\citep[see discussion in Section 1.7.1 of][]{shalchi2009}.
~In the current version model, we use $\rho_i=0.06$ as discussed in \citet{LIS_ApJ2017}.
As discussed in~\citet{Bobik2011ApJ}, we used an enhanced $K_{\perp,\theta}$ in the
polar regions; this enhancement is an implicit way of reducing drift effects by changing the CR intensity gradients significantly~\citep{2013LRSP...10....3P} in order to reproduce the amplitude and rigidity dependence of the latitudinal gradients of GCR differential
intensities for protons (see Sect.~\ref{Heliosph_latid_dep} and, e.g., \citealp{2016SoPh..291.2181V}).
\par
Moving towards the solar maximum the rate of coronal mass ejections (CME) increases leading to a more chaotic structure of the solar magnetic field and to a stronger turbulence regime.~In such conditions QLTs and other theories derived for a weak turbulence are no longer valid.~In particular non linear effects are expected to be stronger for particles scattered at pitch angles around 90$^\circ$\citep{shalchi2009}.~As an example, simulations performed in strong turbulence condition showed a linear rigidity dependence of $K_{||}$ that extends to lower rigidity with respect to the value evaluated with QLTs~\citep[e.g., see Figure 3.5 and 6.5 of][]{shalchi2009}.~The effects of such a modification in the theories regard mainly rigidities lower than those considered in this work.~Therefore, for sake of simplicity, the parametrization implemented in \helmod{} model was that expressed in Eq.~(\ref{EQ::KparActual}), where $g_{low}$ decreases down to zero during high activity periods, while $g_{low}
$ reaches its maximum value ($g_{low}=0.3$) for low solar activities\footnote{In current approach, $g_{low}$ was separately tuned for each set of observation.~The resulting \textit{best} values were then fitted to an empirical function, which, in turn, was tuned similarly to the other \helmod{} parameters, discussed in Sect.~\ref{SEC::Parameters}.~A further discussion on $g_{low}$  values during the transition from low to high activity is available in \citet{LIS_ApJ2017}.\label{fittingFootnote}}.
For sake of completeness, as reported by, e.g.,~\citet{2014JGRA..119.2411G}, merged interaction region (MIR)\footnote{A MIR is the buildup of multiple interplanetary ejecta with enhanced solar wind speed, magnetic field, and plasma density} can additionally reduce the diffusion and drift coefficients~\citep{1999JGR...104.4709L}, although not explicitly treated in this work, the time dependence of diffusion parameter includes such effect as it is linked to the real variation of GCR fluxes.
\par
Finally, it has to be remarked that in the so-called \textit{force-field model} (FFM) (e.g.,~see
\citealt{FFM1968,1971Ap&SS..11..288G}, Section 2.1 of~\citealp{Bobik2011ApJ}, also Section~8.2.4
of~\citealt{rancoita2015} and references therein), the diffusion tensor is reduced to a scalar for spherically symmetric modulated number density of CR particles and steady-state modulation conditions.~The FFM is an approximated way for treating solar modulation in many practical applications, but its intrinsic assumptions do not allow to account for relevant effects, like those related to the charge drift reported, for instance, from the observation of GCR modulation during periods with opposite field polarities of Sun~\citep[for instance, by ][]{EvensonMeyer1984,GarciaMeyer1986,ClemEvenson2000,BoellaEtAl2001}.
Besides, evidences on how drift mechanisms can modify both the radial and (solar) latitude gradients were reported by~\citet[and reference therein]{CummingsetAl1987,McKibben1975,Simpson1996a,Heber1996,Heber2008,deSimone2011,GieselerHeber2016}.~For a further discussion about limitations of the FFM to describe the modulation of cosmic rays, one can see, for instance,~\citet{Caballero2004}.~Nevertheless, it is worth to remark that the time variation of FFM modulation potential can be considered a good time variation proxy for the global behavior of heliosphere and, in turn, it can be exploited for determining the diffusion parameter (e.g., see Section 2.1 of~\citealp{Bobik2011ApJ}, Section~8.2.4
of~\citealt{rancoita2015} and Sect.~\ref{Sec::K0Fit}).

\begin{figure*}
\centering
\includegraphics[width=0.75\textwidth]{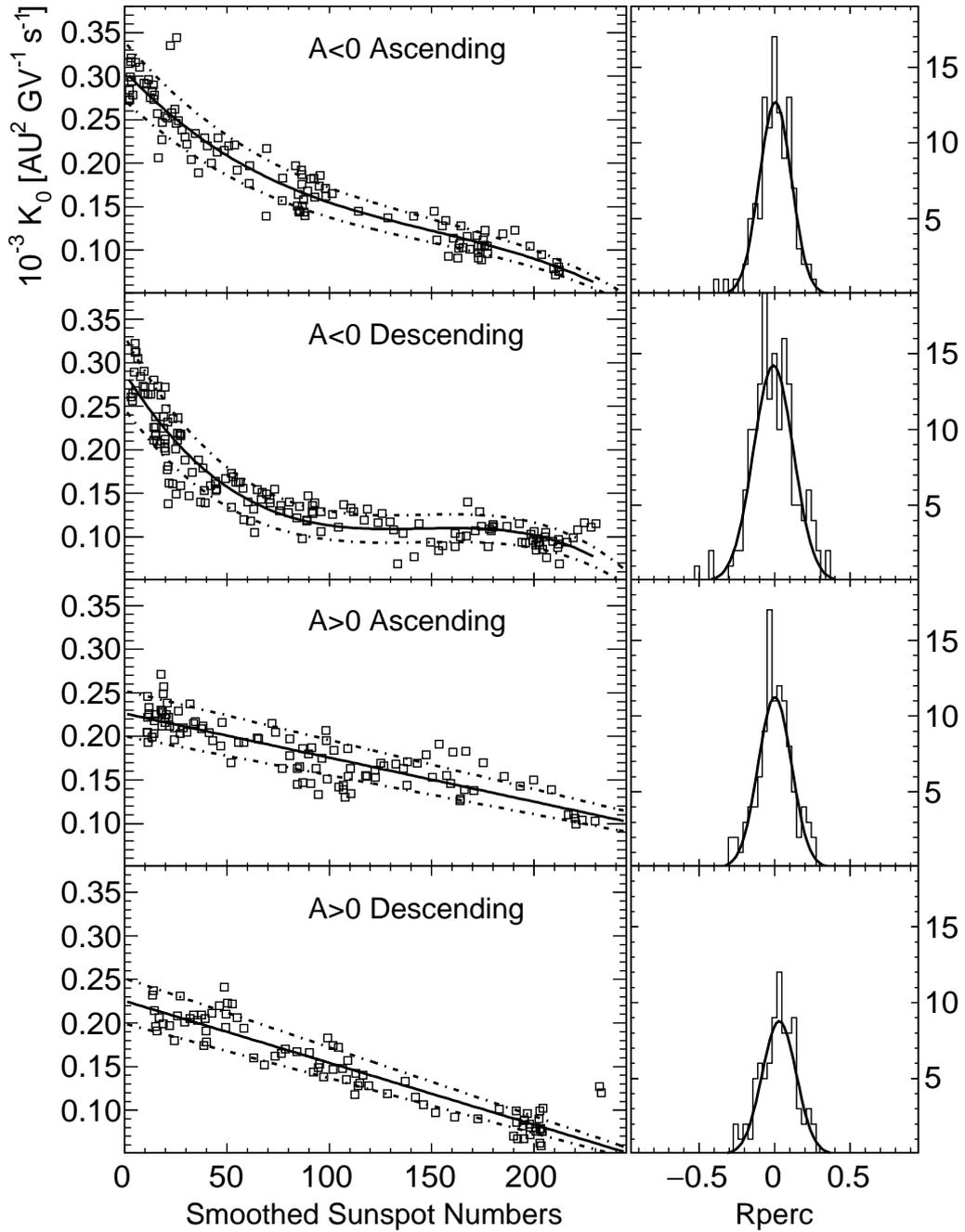}
\caption{Left - Diffusion parameter $K_{0}$  as a function of the SSN value; the
central continuous lines are obtained from a fit of $K_{0}$ with
respect to SSN values in the range $10 \lesssim \textrm{SSN} \lesssim
165$; the dashed and dotted lines are obtained adding (top) or
subtracting (bottom) one standard deviation from the fitted values.  Right - distribution of relative differences $R_{\rm perc}$. }
\label{fig:FitDistro1}
\end{figure*}
\begin{table*}
  \caption{Parameters $c_i$ of the polynomial expression (\ref{eq:cubic})
  as a function of solar polarity and phase. In the last column the rms value of the relative differences is shown. }
  \begin{center}
  \begin{tabular}{ lccccc }
    \hline
		    &$c_0$	&$c_1$		&$c_2$		&$c_3$		& rms\\
    \hline
    A$<$0 Ascending &0.0003059	&-2.51e-6	&1.284e-8	&-2.838e-11	& 0.1097\\
    A$<$0 Descending&0.0002876	&-3.715e-6	&2.534e-8	&-5.689e-11	& 0.1400\\
    A$>$0 Ascending &0.0002262	&-5.058e-7	&--		&--		& 0.1153\\
    A$>$0 Descending&0.0002267	&-7.118e-7	&--		&--		& 0.1607\\
    \hline
  \end{tabular}
\label{table1}
\end{center}
\end{table*}
\subsection{The Diffusion Parameter}
\label{Sec::K0Fit}
%
%
%

\begin{table*}
  \caption{Parameters $p_i$ of the exponential expression (\ref{eq:expo})
  for the high solar activity. In the last column the rms value of the relative differences is shown. }
  \begin{center}
  \begin{tabular}{ lcccc }
    \hline
	  &$p_0$ 	&$p_1$	&$p_2$		& rms\\
    \hline
    MCMU  &0.003753	&-0.04791& 0.0001365	& 0.100\\
    OULU  &0.001354	&-0.10070& 0.0007697	& 0.094\\
    \hline
  \end{tabular}
  \label{table2}
  \end{center}
\end{table*}
\par
The diffusion parameter $K_0$ introduced in Eq.~(\eqref{EQ::KparActual})
(e.g., see Section 2.1 of~\cite{Bobik2011ApJ}) is a scaling factor that defines the global behavior of particle flux modulation in the heliosphere and its dependence on time reflect the variability of interplanetary medium properties (like the actual solar magnetic field transported by SW and its turbulence) during the different phases of solar cycles~\citep[e.g., see Equation 4 in][]{ManuelFerreiraPotgieter2014}.~$K_0$ was expressed in Section 2.1
of~\citealt{Bobik2011ApJ} and afterwards in~\citealt{DellaTorre2013AdvAstro} by means of
a practical relationship with respect to the monthly smoothed
sunspot numbers (SSN); in those papers, such a relationship was demonstrated to be adequate for the
description of the dependence of the diffusion parameter on
solar activity and polarity.
\par
The current $K_0$ employed values are derived by means of the procedure discussed in Section 2.1 of~\cite{Bobik2011ApJ} using the data from \citep{Usoskin2011} and
the SSN, updated to the most recent data series \citep{sidc,ssn_v2}; they
are subdivided into four subsets, that is, ascending and descending
phases for both negative and positive solar magnetic field
polarities and are shown in Fig.~\ref{fig:FitDistro1}.~The updated practical relationship between $K_0$ in ${\rm
AU^2GV^{-1}s^{-1}}$ and SSN values for $2.2\le{\rm SSN}\le266.9$ for
those periods (see Fig.~\ref{fig:FitDistro1}) found is:
\begin{equation}
  K_0^{SSN} = c_0 + c_1 \,{\rm SSN} + c_2 \,{\rm SSN}^2 + c_3 \, {\rm SSN}^3
  \label{eq:cubic}
\end{equation}
with the parameters $c_i$ listed in Table~\ref{table1}.~Furthermore, the root mean square (rms) values of the relative differences between the values
obtained using Eq.~\eqref{eq:cubic} -- following the procedure discussed in \cite{Bobik2011ApJ} --
and current $K_0$ data are also reported in Table~\ref{table1}.~It can be shown that the practical relationship \eqref{eq:cubic}
provides an overall agreement between calculated diffusion parameters, as
function of SSN, and $K_0$ values.~
\par
This description is good enough to deal with CR
modulation with \helmod{} code for periods of low solar activity.~However, as soon as the high solar activity periods are considered
-- that is, in current approach, when tilt angles are $\gtrsim 50^{\circ}$ --, the rate and intensity of disturbances
from Sun, such as CMEs and their short-term impacts, become
increasingly larger, i.e., are resulting in a more chaotic structure with i)
stronger perturbations of the magnetic field large-scale structure, which is hardly describable
by means of a simple dipole approximation, and ii) local modulation effects which can be only related empirically to the actual SSN value.~Therefore, a different or an additional
solar activity indicator should be investigated.~In fact, the monthly
smoothed CR counting rates recorded by \textit{neutron monitors} (NM) allows
to better reproduce the short-time variation of $K_0$ that are needed to
correctly calculate the CR modulation.~ Among the available set of neutron monitor
counting rate (NMCR) data series{~\citep{nmdbWeb,2011AdSpR..47.2210M}}, those from McMurdo NM (MCMU) in Antarctica
(with effective vertical cutoff rigidity $P_c \simeq 0.3$\,GV) and Oulu NM~\citep{1991ICRC....3..145K} in
Finland (with $P_c \simeq 0.8$\,GV) were considered.~These two mentioned NM stations have a low cutoff rigidity with a long enough period of data
taking.~The vertical {geomagnetic} cutoff rigidity is an
adequate approximation of the rigidity lowest-limit of the primary
spectrum to which the NM is sensitive\footnote{This is generally true for geographic position which have a geomagnetic cutoff rigidity below the atmospheric threshold, that for proton $\sim 1$ GV.}.~However, the maximum of NM
sensitivity -- i.e. the maximum of the response function~\citep{ClemDorman2000} -- with respect to interactions of primary cosmic ray particles with the atmosphere, occurs 
in the rigidity interval 3--15\,GV, depending on the NM site~\cite[e.g., see][for a further discussion]{JGRA:JGRA52380}.~Using monthly NMCR data, the new practical relationship for the high
solar activity periods becomes:
\begin{equation}
  K_0^{NMCR} = p_0~ \exp\left( p_1 \,{\rm NMCR} + p_2\, {\rm NMCR}^2 \right)
  \label{eq:expo}
\end{equation}
with the parameters $p_i$ listed in Tab.~\ref{table2}.
For sake of completeness, we investigated also the usage of NMCR data for low activity periods.~For such a purpose, a fit to a relationship similar to that provided by Eq.~\eqref{eq:expo} was performed.~It can be shown that the so-obtained diffusion parameters allow to derive modulated spectra similar to those in which $K_0$ is obtained by means of Eq.~\eqref{eq:cubic}.~Therefore, since the use of NMCR during low solar activity does not result in an appreciable difference, we kept the approach employing SSNs, as activity indicator for such periods.
\par
{It has to be noted that, although those two stations have different sensitivities and values of vertical rigidity cutoff,
in a relative scale the relationship
between NMCRs and the diffusion parameter is similar}.~This can be shown, for example, by
computing the rms's of the relative differences between values
obtained using Eq.~\eqref{eq:expo} and $K_0$ data.~For the two NM stations considered the rms values found are shown in Tab.~\ref{table2} for MCMU and OULU, respectively.~The found results are compatible and, thus, only the relation \eqref{eq:expo} with MCMU data (shown in Fig.~\ref{fig:FitDistro_high}) was employed in this work,
since it refers to the station with the lowest vertical cutoff.
\begin{figure}[htb]
\centering
\includegraphics[width=0.48\textwidth]{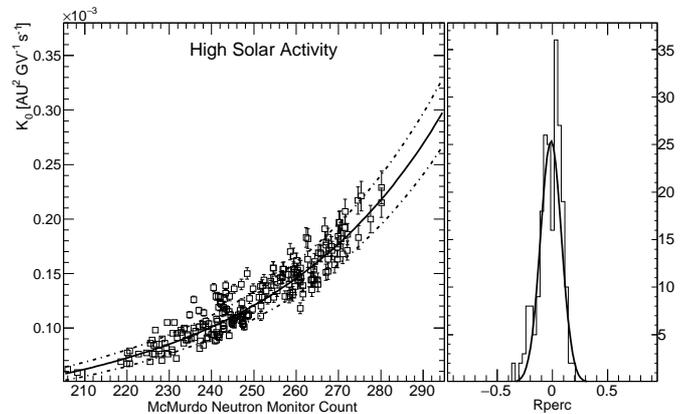}
\caption{Left - Diffusion parameter $K_{0}$
as a function of the McMurdo Neutron
Monitor Counting rate during periods of high solar activity; the
central continuous lines are obtained from a fit of $K_{0}$ with
respect to NMC values in the range $210 \lesssim \textrm{NMC} \lesssim
300$; the dashed and dotted lines are obtained adding (top) or
subtracting (bottom) one standard deviation from the fitted values. Right - distribution of relative differences $R_{\rm perc}$. }
\label{fig:FitDistro_high}
\end{figure}
\subsection{The Interplanetary Magnetic Field}
\label{SEC::HMF}
The GCR propagation in the heliosphere is affected by the outwards flowing SW
with its embedded magnetic-field and magnetic-field irregularities.~The Solar magnetic field is transported by the non-relativistic streaming particles of the SW, which carries the field into interplanetary
space, producing the large-scale structure of the interplanetary (or heliospheric) magnetic field and the heliosphere geometry.~
The so-generated and transported IMF is characterized
by both the large scale structure (SW expansion from a rotating source) and
low scale irregularities that change with time according to solar activity
(e.g., due to variations of SW velocity or to plasma perturbations related to CMEs).~
\par
In the current version of \helmod{} we used IMF and drift as implemented in~\citet{DellaTorre2013AdvAstro}.~The heliosphere is divided into \textit{polar} and \textit{equatorial} regions, where different descriptions of IMF are applied.~In the \textit{equatorial} region, we used the Parker's IMF ($\vec B_{Par}$) in the parametrization of~\citet{Hattingh1995},
while in the \textit{polar} regions we used a modified IMF ($\vec B_{Pol}$) that includes a latitudinal component,
accounting for large scale fluctuations, dominant at high heliolatitudes, as suggested by~\citet{JokipiiKota89}:
\begin{eqnarray}
\vec B_{Par} &=& \dfrac{A}{r^{2}}\left(\vec e_{r}-\Gamma\vec e_{\varphi}\right)\left[1-2H\left(\vartheta-\vartheta'\right)\right]\label{IMFparpol_Par}\\
\vec B_{Pol} &=& \vec B_{Par}+ \dfrac{A}{r^{2}}\left[ \dfrac{r}{r_b}\delta(\vartheta)\vec e_{\vartheta} \right] , \label{IMFparpol}
\end{eqnarray}
with
\[
 \Gamma=\frac{\Omega (r-r_0) \sin\vartheta}{V_{sw}}.
\]
In Eqs.~(\ref{IMFparpol_Par}, \ref{IMFparpol}),
$A$ is a coefficient that determines the IMF polarity and allows
$|\vec B_{Par} |$ to be equal to $B_\oplus$, i.e.,
the value of the IMF at Earth's orbit -- as extracted from NASA/GSFC's OMNI data set through OMNIWeb
\citep{SW_web};  $\vec e_r $, $\vec e_\theta $ and $\vec e_\varphi $ are unit vector components in the
radial, latitudinal  and azimuthal directions, respectively; $\vartheta$ is the co-latitude (polar angle);
$\vartheta'$ is the polar angle determining the position of
the heliospheric current sheet (HCS); $H$ is the Heaviside function: thus, the term, $[1-2H(\vartheta-\vartheta')]$ allows $\vec B_{Par}$
to change sign in the two regions above and below the HCS~\citep{1981ApJ...243.1115J} of the heliosphere.~Furthermore,
$\Omega$ is the angular solar rotation speed and is assumed to be independent on the heliographic latitude and equal to the
sidereal rotation at Sun's equator.~The separation between \textit{equatorial} and \textit{polar} regions were set to $\theta=30\degree$ and $\theta=150\degree$ according to~\citet{DellaTorre2013AdvAstro}.~Finally, in order to have a divergence free magnetic field, we require that the perturbation factor [$\delta(\vartheta)$] must be~\citep[and reference therein]{DellaTorre2013AdvAstro}:
\[
\delta(\vartheta)=\frac{\delta_{m}}{\sin\vartheta},
\]
where $\delta_{m}$ is the minimum perturbation factor of the field. The perturbation parameter is let to grow with decreasing
co-latitude, while in their original work~\citet{JokipiiKota89} fixed the value of
$\delta$ between $10^{-3}$ and $3\times10^{-3}$.~In this work we use $\delta_{m}=2\times10^{-5}$, tuned by comparison with Earth orbit observations during solar cycles 23-24~\citep{LIS_ApJ2017}.
{One has to note that the modified IMF introduced an additional magnetic component in the latitudinal direction that was not present in the pure Parker field. This should be considered when the diffusion tensor is generalized to heliocentric coordinates system using relationship reported e.g. in~\citet[and reference there-in]{burg2008}.}
\par
As already reported in the review by~\citet{Owens2013}, while some basic descriptions -- such as the Parker spiral \citep{parker58} -- are
fully developed and already included into standard textbooks on space physics, other topics are still under development at the time of writing.~A correct description of IMF is fundamental in a proper description of GCRs propagation{~\citep[see also][for a discussion about modified Parker's magnetic field]{Raath2016}}.~As matter of fact, observations performed out-of-ecliptic by instruments on-board of Ulysses spacecraft (e.g., see~\citealp{Sandersonetal1995,Marsden2001,BaloghetAl2001})
showed the limit of the Parker field approach towards the polar regions and opened the way to a more complex description~\citep[e.g., see][]{JokipiiKota89,smithbieber1991,Fisk1971,2010AdSpR..45...18H,burg2008}.~Of great importance in the IMF description is the heliospheric current sheet (HCS) that divide the heliosphere
in regions of inward or outward directed magnetic field lines~\citep[e.g., see section 9.1 of ][]{2006RPPh...69..563S}.~The IMF is not a stable structure, but evolves with time following a 11-year cycle usually defined using sunspot numbers~\citep{2015LRSP...12....4H}.~At each cycle, the IMF reverses the magnetic polarity, usually defined by the sign of polar magnetic field strength~\citep[e.g., see][]{1978SoPh...58..225S},
thus the \textit{Hale cycle} or \textit{magnetic cycle} of the IMF lasts $\approx 22$ years.
%
\section{Effective Heliosphere Parameters and LIS's} \label{SEC::Parameters}
%
\begin{table*}
  \caption{Parameters of the analytical fits to the proton and He LIS's (from Table 5 of \citet{LIS_ApJ2017}. }
  \begin{center}
  \footnotesize
  \begin{tabular}{crrrrrrrrrrrrrrr}
    \hline
	  &
{$a_0$} &
{$a_1$} &
{$a_2$} &
{$a_3$} &
{$a_4$} &
{$a_5$} &
{$b$} &
{$c$} &
{$d_1$} &
{$d_2$} &
{$e_1$} &
{$e_2$} &
{$f_1$} &
{$f_2$} &
{$g$} \\
    \hline
    p &
$94.1$ & 
$-831$ & 
0 &
$16700$ & 
$-10200$ & 
0 &
$10800$ & 
$8590$ & 
$-4230000$ & 
$3190$ & 
$274000$ & 
$17.4$ & 
$-39400$ & 
$0.464$ & 
0
\\

He &
$1.14$ &
0 &
$-118$ & 
$578$ & 
0 &
$-87$ & 
$3120$ & 
$-5530$ & 
$3370$ & 
$1.29$ &
$134000$ & 
$88.5$ & 
$-1170000$ & 
$861$ & 
$0.03$  
\\
    \hline
  \end{tabular}
  \normalsize
\label{tbl-4}
\end{center}
\end{table*}
%
\begin{figure}[tb]
\begin{center}
 \includegraphics[width=0.45\textwidth]{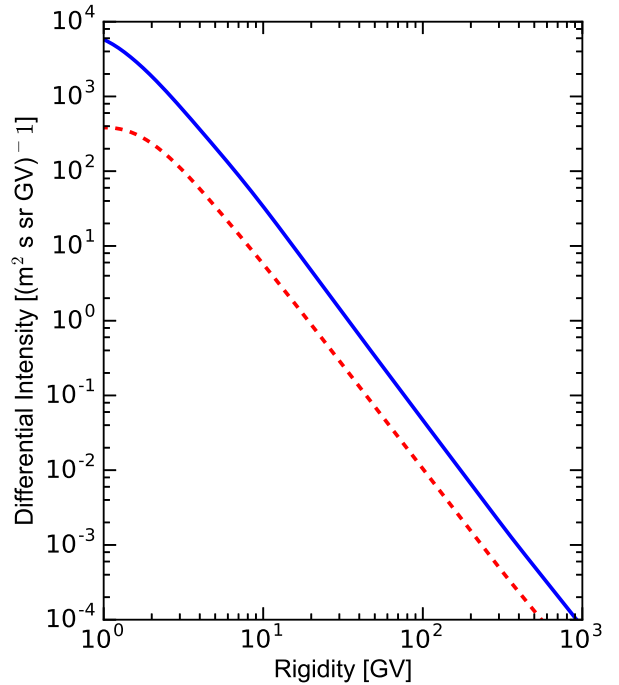}
 \caption{Local interstellar spectra as obtained by \citet{ECRS2016b,LIS_ApJ2017} for CR protons (blue solid line) and helium (red dashed line) above 1\,GV.}
 \label{fig:GALPROPLISS}
\end{center}
\end{figure}
One of the success of \helmod{} model with respect to other available Monte Carlo codes for heliospheric propagation~ -- like, e.g., \textit{SolarProp}--\citealp{2014PhRvD..89h3007G} and \textit{HelioProp}--\citealp{2016CoPhC.207..386K})-- is its reduced number of free parameters necessary for the description of modulation mechanisms.~These parameters -- as described later along the present section -- are related to quantities determined from observations and need, in turn, to be tuned in order to obtain a comprehensive set of modulated spectra, like those discussed in Sect.~\ref{Results}.
{Free parameters tuned in this work are: a) the ratio between parallel and perpendicular diffusion coefficients and b) the $g_{low}$ parameter (both presented in Sect.~\ref{SEC::kpar}). $\rho_i$ modifies the absolute scale of modulation intensity up to high rigidities; since this parameter also influences latitudinal gradients, its value is constrained in such a way that the obtained latitudinal gradients are in agreement with those found by Ulysses and presented in  Sect.~\ref{Heliosph_latid_dep}.
The maximum value of $g_{low}$ tunes modulated differential intensity in the low rigidity range ($<$ 3 GV), mainly during low activity periods. In present approach other parameters, i.e., $\delta_m$ (Sect.~\ref{SEC::HMF}) and \textit{drift suppression factor}(Sect.~\ref{SEC:parkerEq}) where not additionally tuned with respect to those previously estimated~\citep[e.g., see][]{DellaTorre2013AdvAstro}, since their variation has a minor impact in the considered rigidity range for \helmod{} ($> $1 GV). Furthermore,
the values of the diffusion parameter (described in Sect.~\ref{Sec::K0Fit}) were evaluated using published data up to the end of solar cycle 23, and applied along the full solar cycle 24. However, one has to note that during solar cycle 22 for protons at low solar activity and positive polarity of the solar magnetic field, Ulysses probe has shown the presence latitudinal gradients with respect to the ecliptic plane. These gradients might affect the actual value of $K_0$ which is applied, mainly, as an overall diffusive scale factor for the whole heliosphere, although being evaluated on the ecliptic plane only. }
\par
Since any perturbation caused by Sun propagates into the heliosphere carried by SW, it takes typically $\sim 15$ Carrington rotations to reach the heliosphere boundary.~In the meanwhile, the properties transported by SW change, e.g., new perturbations arise from Sun and propagate outwards, thus creating local areas, within the solar cavity, which have to be differently described.~Therefore, in \helmod{} model the heliosphere is described as an \textit{effective heliosphere}~\citep{Bobik2011ApJ} with a radius of 100\,AU and is subdivided in 15 radially equally-spaced regions.~Thus, in present bi-dimensional approximation, the model assumes that the propagation of GCR occurs within a static and spherical heliosphere with the TS located at 100\,AU (for a further discussion see \citep{Bobik2011ApJ}).~Each \textit{i}-th region experienced by CR particles is characterized by the heliospheric parameters evaluated at \textit{i}-Carrington rotations back-in-time, corresponding to the time necessary to SW to reach
it.~Furthermore, as
discussed in~\citet{ECRS2016b,LIS_ApJ2017}, \helmod{} describes modifications of heliosphere dimensions as changes affecting only the \textit{effective distance} between Earth and Sun.~For example, in the re-scaled \helmod{} effective heliosphere if -- because of the variation of SW ram pressure -- TS is moved by 10\,AU \citep[e.g., see][]{WashimiEtAl2011}, Earth's \textit{effective location} is only shifted by 0.1\,AU; Monte Carlo simulations (see Sect.~\ref{Code}) indicate that such variations have negligible effects on the modulated spectra when, instead, are determined at 1\,AU (i.e., at Earth), thus for the purposes of the current work, we can properly assume that real and effective Earth locations are both at 1\,AU from Sun.
\par
It is worthwhile  to remark that $\approx 100\,$AU is the average value of TS locations,
which can be obtained from Table~2 of~\citet{2003ApJ...589..635W}.~Furthermore~\citep[e.g.,~see][]{Stone2005,2008Natur.454...71S},
Voyager 1 and 2 reached the TS in 2004 and 2007 at about 94.0\,AU and 83.7\,AU, respectively,
from Sun, in agreement with the predictions from Whang and collaborators.~\citet{2005ApJ...630.1114L} treated symmetric
and asymmetric TS models and concluded that for $A>0$ cycles at solar minimum no significant difference occurs;
for $A<0$ cycles variations remain negligible in nose direction while, approaching the tail direction,
some differences can be appreciated at proton energies below 1--1.5\,GeV.~However,
\citet{2005ApJ...630.1114L} and \citet{Potgieter_2008} suggested that, in general,
a symmetric TS with a radial distance of $\approx 100\,$AU is still a reasonable assumption.~In addition, it should be noted that, in literature, the heliospheric structure is considered latitudinally asymmetric (particularly) during solar minimum conditions mostly because the SW speed depends on the latitude and solar activity~\citep[e.g., see][]{1998GeoRL..25....1M}.~A more complete asymmetrical structure of the heliosphere can be evaluated using magneto-hydrodynamic (MHD) models~\citep[e.g., see][]{2009ApJ...701..642F,2014JGRA..119.2411G} that include transport
description in outer heliosphere, heliosheath and co-rotating interaction regions.~Moreover, Voyager 1, 2 observations point to a dynamic TS that is moving inward/outward in the heliosphere \citep{Stone2005,RichardsonWang2011},
while numerical models indicate that this TS movement could be as large as $\sim$20 AU over a complete solar cycle \citep[see the discussion in][and references therein]{ManuelFerreiraPotgeter2015}.~Finally,
even though variations of the real size of the heliosphere may be important for the analysis of CR propagation near the TS, we do not consider them in this work.
\par
The diffusion parameter $K_0$ is determined (as discussed in Sect.~\ref{Sec::K0Fit})
using the values of modulation strength, SSN\footnote{http://www.sidc.oma.be/sunspot-data/} values,
NMCR\footnote{http://www.nmdb.eu/nest} values at McMurdo station
and the effective heliosphere radius (see discussion above).~Other parameters (which depend on the solar activity) are
the tilt angle $\alpha_{\rm t}$ of HCS, the
magnetic field polarity [related to the sign of the coefficient $A$ in Eq.~\eqref{IMFparpol}], the
magnetic field amplitude ($B_\oplus$) and, finally, the solar wind velocity ($V_{\rm sw}$).~The latter two parameters are measured at Earth's orbit and provided by OMNIWeb\footnote{http://omniweb.gsfc.nasa.gov/}.~The polarity of the magnetic field (obtained from Wilcox Solar Observatory Polar Field Observations\footnote{http://wso.stanford.edu/}) and $B_\oplus$ determine the IMF described by means of Eq.~\eqref{IMFparpol_Par}.~In the current model we use the so-called  ``line-of-sight'' model (L-model) for the tilt angle $\alpha_{\rm t}$ of HCS~\citep{Hoeksema1995}, that provides an overall general agreement with experimental data~\citep[e.g., see discussion in][]{Bobik2011ApJ,DellaTorre2013AdvAstro,ECRS2016b,LIS_ApJ2017}.~$\alpha_{\rm t}$ and the field polarity, obtained from Wilcox Solar Observatory, are used to deal with the drift velocity (as discussed in Sect.~\ref{SEC::HMF}),
which contributes to the overall convection velocity in Eq.~(\ref{eq_parker}).~Drift contribution is relevant during low solar activity -- e.g., for $\alpha_{\rm t} <30^\circ$  --
and decreases with increasing solar activity.~Since during the high activity period the heliospheric magnetic field is far from
being considered regular, in analogy with other works~\citep[see e.g.][]{Potgieter_2008}, we introduced a correction factor that suppress any drift velocity at solar maximum.~Finally, the latitudinal dependence of  the SW speed is the one described in~\citep{Bobik2011ApJ}.
\par
\par
The local interstellar spectra (LIS) are input cosmic ray intensities for any modulation models.~Fluxes are assumed isotropically distributed at heliosphere boundary, in a steady-state configuration.~Recently, \citet{ECRS2016b,LIS_ApJ2017} deduced LIS's for protons, helium and antiprotons using the most recent experimental results combined with the state-of-the-art models (i.e., \galprop{}\footnote{http://galprop.stanford.edu} and \helmod{}) for propagation in galaxy and heliosphere.~In fact,
\helmod{} and \galprop{}~\citep{1998ApJ...493..694M,1998ApJ...509..212S} were combined to provide a single framework and run to reproduce a comprehensive set of observations of CR species collected in different time periods, from 1997 up to 2015.~The authors proposed an analytical expression  
for proton and helium nuclei~\citep{ECRS2016b,LIS_ApJ2017} LIS's as a function of the rigidity expressed in unit of [m$^2$ s sr GV]$^{-1}$ (see also Fig.~\ref{fig:GALPROPLISS}):
\begin{align}\label{EQ::an}
 &J_{LIS}(P)\times P^{2.7} = \\
&\left\{
\begin{array}{ll}
\sum_{i=0}^5 a_i P^{i}, &P\le1\ {\rm GV},\nonumber\smallskip\\
b + \frac{c}{P} + \frac{d_1}{d_2+P} + \frac{e_1}{e_2+P} + \frac{f_1}{f_2+P} + g P, & P>1 \ {\rm GV},\nonumber
\end{array}
\right.\nonumber
\end{align}
where $a_i, b, c, d_i, e_i, f_i, g$ are the numerical coefficients summarized in Table~\ref{tbl-4} (from Table 5 of \citet{LIS_ApJ2017}).
\section{The Monte Carlo Code}
\label{Code}
For most applications Parker's transport equation [Eqs.~\ref{eq_parker} and \ref{eq_parker_p}] has been solved using numerical methods, because its intrinsic complexity.~
\par
The traditional approach to solve multi-dimensional partial differential equations makes use of numerical integration methods such as the finite difference technique~\citep[e.g., see][]{JokipiiKopriva1979,KotaJokipii1983,Potgieter85,BurgerHatting1995} or
as the standard implicit difference technique~\citep[e.g., see][]{Fisk1971,KotaJokipii1991}.~These methods have several disadvantages, mainly numerical instability
problems when solving differential equation in higher dimensions~\citep{PeiBurger2010,Kopp2012}.~A modern approach that in recent years  has been
used more frequently to solve numerically a variety of problems in space-physics is based on
Monte Carlo methods~\citep[e.g., see][]{kruell1994,FichtnerEtAl1996,Yamada1998,GervasiEtAl1999,Zhang1999,alanko2007,PeiBurger2010,Strauss2011,Bobik2011ApJ,2014PhRvD..89h3007G,2016CoPhC.207..386K}.~A diffusion process described by a FPE can be written as well with a set of stochastic differential equations (SDE)~\citep[e.g., see Chapter 1.6-1.7 of][]{kloeden1999}.~As reported in~\citet[and reference therein]{DellaTorre2016_OneD}
this approach allows to get more flexibility in model
implementation, more stability of numerical results and more possibility to explore physical results that are hard
to handle with  ``classical'' numerical methods.~With the stochastic approach, the solution can be
evaluated computing the SDEs both ``forward-in-time'' or ``backward-in-time''.~A comparison between the two approaches with a numerical estimation of systematic uncertainties can be found in~\citet{DellaTorre2016_OneD}.~In ``forward-in-time''  approach quasi-particle objects were traced from the heliosphere boundary down
to the inner part of heliosphere.~In ``backward-in-time'' approach the numerical process starts from the
target and trace-back quasi-particle objects till the heliosphere boundary.~The ``backward-in-time'' method is widely used, due to faster evaluation of spectra at single points inside
the heliosphere and is presented in this section.~
\par
A stochastic motion does not allow for a single particle study, but it is only possible to
explore how the system evolves in average, considering all the particles as an ensemble.~When the particles
stochastic behavior is studied from probabilistic point of
view, the theory of Markov stochastic process may provide
very powerful mathematical tools~\citep{Zhang1999}.~As pointed out by~\citet{PeiBurger2010}, it is interesting to note that,
since each random process have to be independent of all others, one major
advantage of the stochastic method is that it is
very easy to parallelize the computation.~Therefore, it can run in the same time on many
CPUs of a local cluster reducing significantly the computation time and the hardware costs.
\par
The equivalent set of SDEs, for the 2D approximation, is derived from the 3-D transport equations (reported in~\ref{app1}) by integrating over the azimuthal component:
\begin{subequations}\label{eq::SDE_Hel2}
\begin{align}
 \Delta r =&  \left[\frac{1}{r^2}\frac{\partial r^2K_{rr}}{\partial r}
                    +\frac{1}{r\sin\theta}\frac{\partial}{\partial \theta}\left( K_{\theta r}\sin\theta\right)
                     \right]\Delta t \nonumber \\
         &   - \left[V_{SW} + v_{drift,r}\right]\Delta t \nonumber \\
         &   + (2K_{rr})^{1/2} \omega_{r} \sqrt{\Delta t},\\
\Delta \theta =&  \left[
                    \frac{1}{r^2}\frac{\partial rK_{r\theta}}{\partial r}
                  + \frac{1}{r\sin\theta}\frac{\partial}{\partial\theta}\left(\frac{K_{\theta \theta}\sin\theta}{r}\right)
                  - \frac{V_{drift,\theta}}{r}
                   \right]\Delta t \nonumber \\
               & + \frac{ 2 K_{r\theta}}{r\sqrt{2K_{rr}}}  \omega_{r} \sqrt{\Delta t}\\
               &   + \left[ \frac{2 K_{\theta\theta}}{r^2} - \frac{ 2 K^2_{r\theta}}{r^2 K_{rr}} \right]^{1/2}  \omega_{\theta} \sqrt{\Delta t},\\
\Delta T =&  \frac{\alpha_{\rm rel} T V_{SW}}{3 r} \Delta t, \label{eq::SDE_Hel2_deltaT}\\
L =& \frac{2V_{sw}}{r} \left (\frac{1}{3}\frac{\partial \alpha_{\rm rel} T}{\partial T} -1 \right),
\end{align}
\end{subequations}
where $K_{ij}$ is the symmetric part of diffusion tensor in heliocentric spherical coordinates,
$v_{drift,i}$ follows the definition of Eq.~\eqref{drift_antisym_ten},
$V_{\textrm{sw}}$ is the magnitude of solar wind speed and
$\omega_i$ is a random number following a Gaussian distribution with zero average  and unit
standard deviation.~For stochastic integration with $L\neq 0$ the diffusion/convection process alone is not able to
proper describe the stochastic evolution.~Furthermore, $L$ has the meaning of an additional
sources and losses term in FPE (see~\ref{app1}) distributed inside the phase-space, such that
it contributes to the solution with exponential corrections integrated along the stochastic path~\citep{DellaTorre2016_OneD}.~
\par
The vector $\vec q = [r,\theta, T ]$ represents a phase space density element, that in literature is usually labeled as a pseudo-particle, which time evolution is simulated by means of  Eq.~\eqref{eq::SDE_Hel2} from the inner part heliosphere up to the outer boundary.~As presented in~\citet[and reference therein]{DellaTorre2016_OneD}
the differential intensity $J$ can be obtained from the
density of pseudo-particles by averaging over many realizations of the SDEs.
\par
The procedure used to integrate the SDEs \textit{backward-in-time} and to evaluate the solution at Earth (at $r=1$\,AU and $\theta=\frac{\pi}{2}$) is described in Section 4.1.2 of~\citet[and reference therein]{DellaTorre2016_OneD}, i.e.,\\
(a) pseudo-particles are generated at Earth with initial Energy $T_i$;\\
(b) each event is integrated over the time evolution of a pseudo-particle following the path described in Eq.~\eqref{eq::SDE_Hel2},
the integrated ``loss'' term ($L_{Int}$) is incremented by the quantity $L \Delta t$;\\
(c) when a pseudo-particle reaches the outer border of the effective heliosphere located at 100 AU ($r_b$) the value of differential intensity at boundary for the reconstructed energy $T_r$ is saved and, then, weighted for the ``loss'' factor $\exp{(-L_{Int})}$;\\
(d) the modulated differential intensity at Earth for initial energy $T_i$ is evaluated over $N$ realizations by mean of Equation 22 in~\citet{DellaTorre2016_OneD}:
\begin{equation}
J_{Earth}(T_i) = \frac{\beta(T_i)}{N}\sum_{k=1}^N \frac{J_{LIS}(T_{{r},k})}{\beta(T_{{r},k})}\cdot \exp( -L_{Int,k}).\label{eq::JbackT}
\end{equation}
Equations~(\ref{eq::SDE_Hel2}, \ref{eq::JbackT}) allow one to evaluate modulated spectrum as solution of Eq.~\eqref{eq_parker}.~The differential intensity evaluated using Eq.~\eqref{eq_parker_p} follows the same procedure here described taking into account the relationship reported in Eq.~\eqref{eq::DiffIntensity}.~In this case the pseudo particle evolves in a phase-space that includes particle momentum instead of kinetic energy per nucleon (see~\ref{app1} for details on how to derive 2-D SDE equations from generic 3-D solution).
\par
An alternative approach is to use Monte Carlo integration to evaluate the normalized probability
function $G(P_0|P)$ that gives a probability for a particle to be observed at Earth with a rigidity $P_0$ having a rigidity $P$ at the heliospheric boundary.~Once $G(P_0|P)$ is evaluated it is possible to obtain the modulated spectrum directly from an arbitrary $J_{\rm LIS}$ provided, e.g., by \galprop.~
The modulated spectrum at specific rigidity $P_0$ is, then, obtained by~\citep[e.g., see][]{PeiBurger2010,ECRS2016a,LIS_ApJ2017}:
\begin{equation}\label{eq::PyMod_modulation}
 J_{\rm Earth}(P_0)= \int_0^\infty J_{\rm LIS}(P)G(P_0|P)dP.
\end{equation}
This approach allows one to reduce the amount of simulations, when testing several LIS's.~Actually, in previous approach, a new LIS meant Monte Carlo realizations to be re-run for the same heliosphere parameters, while the latter approach allows one to use the same simulations for all LIS's under test.~We implemented an on-line calculator\footnote{http://www.helmod.org} which, using a python script, reads the \galprop{} outputs and provides the modulated spectra for periods of selected experiments for comparison with published data.~The calculation of heliospheric propagation is replaced by the integration of Eq.~(\ref{eq::PyMod_modulation}) using the normalized probability functions, which are pre-evaluated by the \helmod{} code as described in the previous section.~This method dramatically accelerates the modulation calculations while provides the same accuracy of the full-scale simulation.
\section{Comparison with Observations During Solar Cycles 23-24}
\label{Results}
The current \helmod{} model provided modulated differential intensity for protons, helium nuclei and antiproton for low and high solar activities (as discussed in Sect.~\ref{Introduction}).~In this article, we focus on \helmod{} results regarding protons and helium, whose LIS's were recently investigated in~\citet{ECRS2016b,LIS_ApJ2017}.
The current parametrization is also suited  to reproduce the high energy behavior of the measured spectra (e.g., see Figures 3 and 5 in~\citealt{ECRS2016b}
regarding data up to 1 TV).
\subsection{Low Solar Activity}\label{Sect::LowSolar}
%
\begin{figure}
\centerline{
 \includegraphics[width=0.485\textwidth]{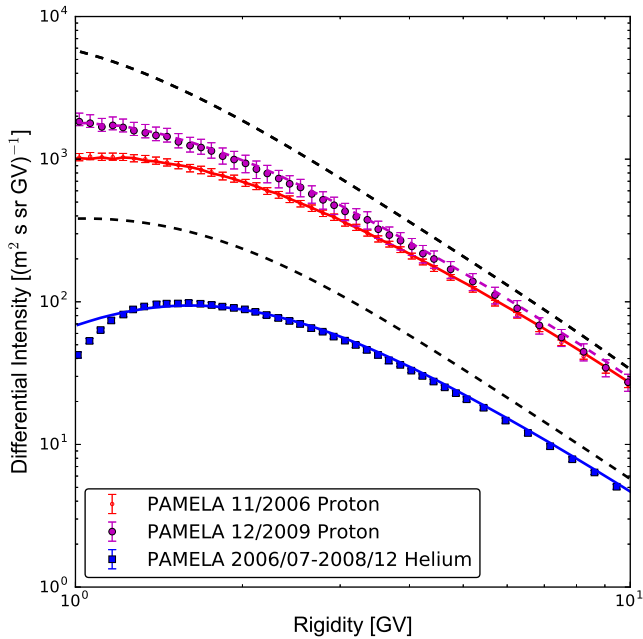}
}
 \caption{Differential intensity of galactic proton~\citep{PamelaProt2013} and helium nuclei~\citep{PAMELA_Prot_He_2011} measured by PAMELA compared with modulated spectra from \helmod{}; the dashed lines are the \galprop{} LIS's  (see text). Comparison of Simulations and experimental data up to 1 TV can be found in~\citet{ECRS2016b}. }
 \label{fig:ProtonHelium_Low}
\end{figure}
\begin{figure*}[htbp]
\begin{center}
 \includegraphics[width=0.8\textwidth]{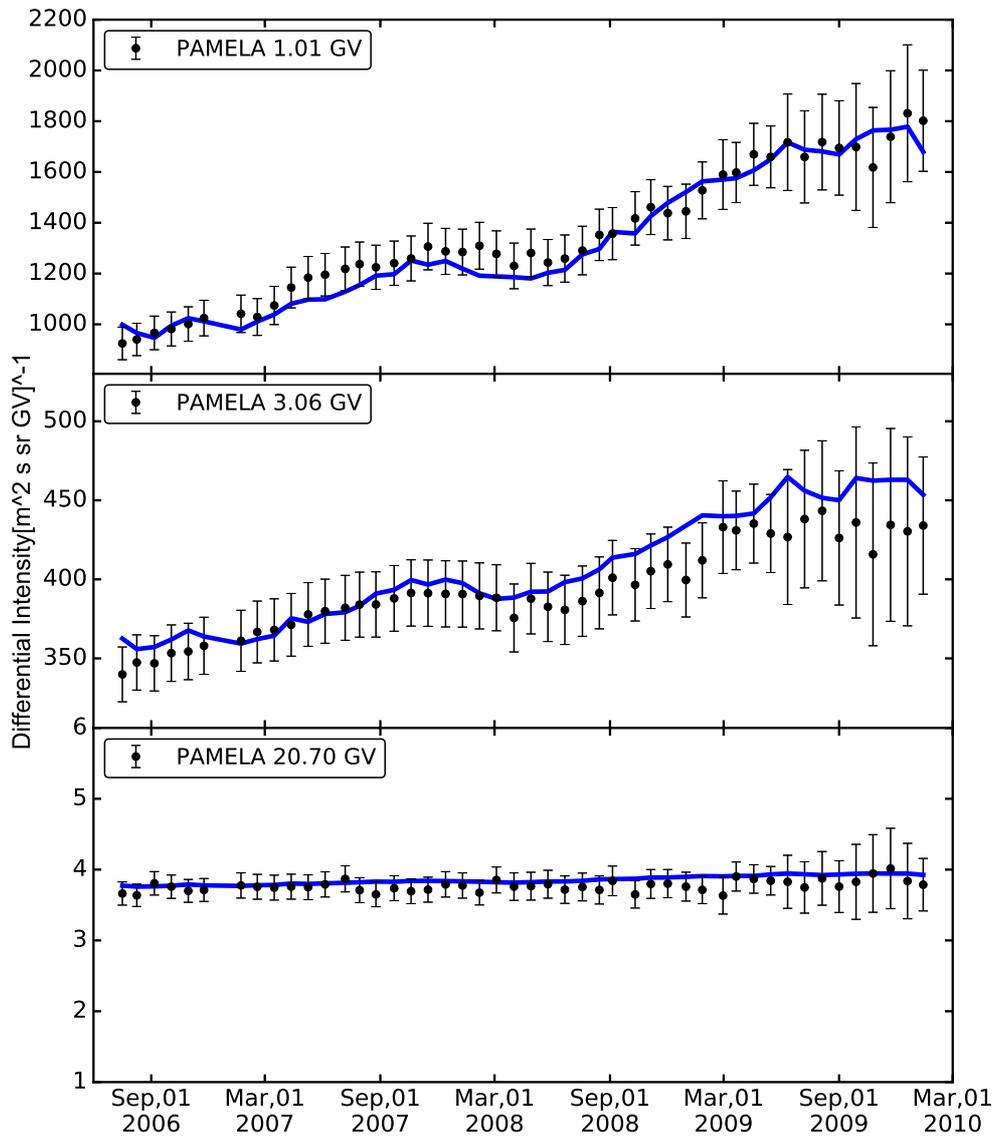}
\end{center}
 \caption{Differential intensity measured by PAMELA for galactic protons\citep{PamelaProt2013} in three different rigidity bins compared with modulated intensities from \helmod{} (see text).}
 \label{fig:PAMELATime_Low}
\end{figure*}
\begin{figure*}[htb]
\centerline{
 \includegraphics[width=0.8\textwidth]{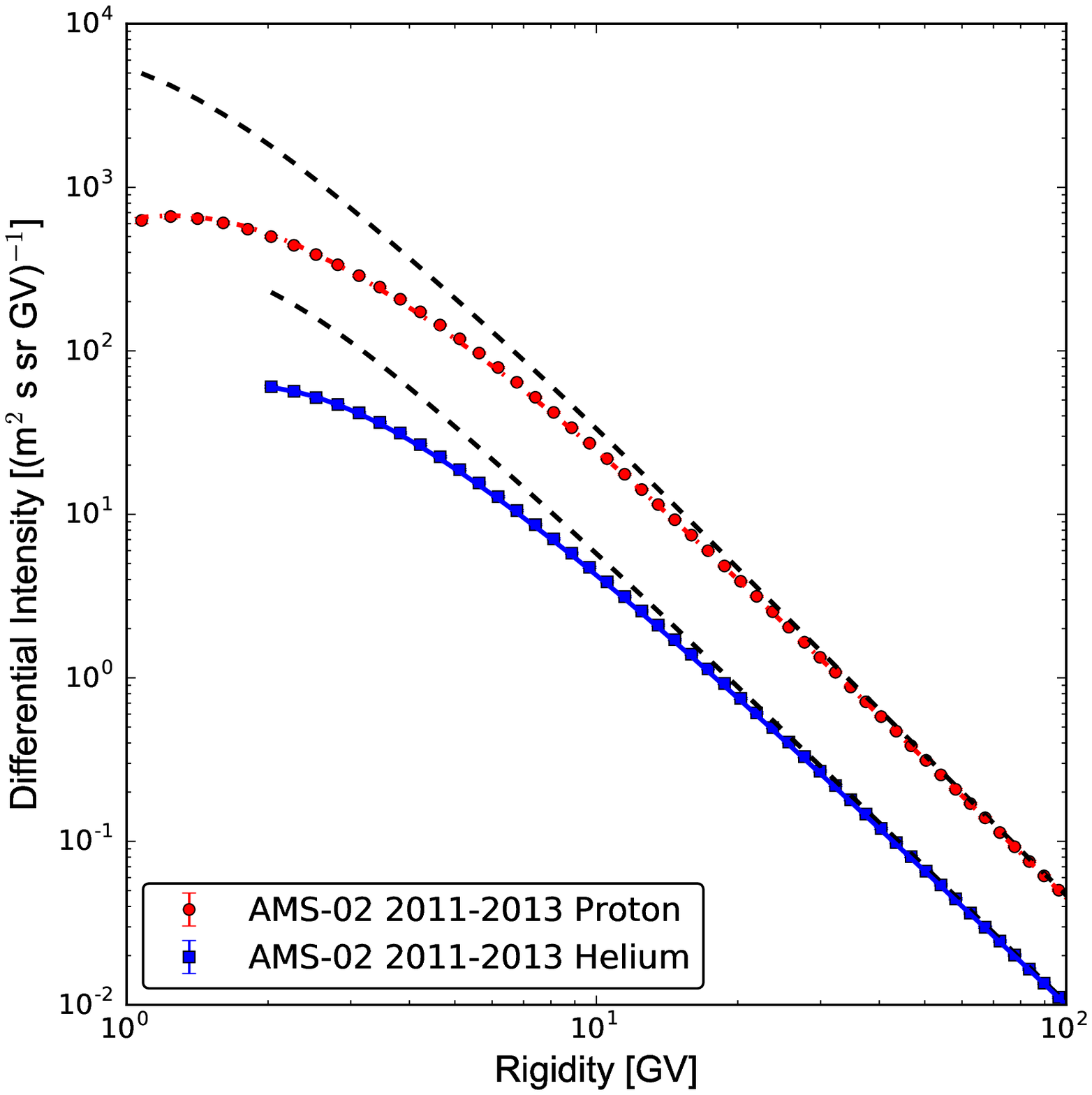}
 }
 \caption{AMS-02 differential intensities for galactic protons\citep{AMS02_2015_PhysRevLett1} and helium nuclei~\citep{AMS02_2015_PhysRevLett2}.~Points represent the experimental measurements, the dashed lines is the \galprop{}
 LIS and solid lines are the computed modulated spectra from \helmod{} (see text). Comparison of Simulations and experimental data up to 1 TV can be found in~\citet{ECRS2016b}.}
 \label{fig:pHe_High}
\end{figure*}
\helmod{} was initially developed for periods of low solar activity, when the contribution of charge dependent transport processes -- resulting in a magnetic drift convection, i.e., that expressed by $\vec{v}_{drift}$ in Eq.~\eqref{eq_parker} -- is so important enough to introduce peculiar features, for instance, in the time dependence observed for the positron fraction~\citep[e.g., see][]{DellaTorre2012} at rigidities lower than (10-20)\,GV and heliospheric latitudinal gradients of galactic cosmic rays distribution~\citep[e.g., see][]{DellaTorre2013AdvAstro}.
\par
The large-scale structure of the IMF is strongly affected by solar activity.~More the IMF assumes a regular structure, more GCR particles experience the effects of magnetic drift transport.~As already mentioned, the relevance of magnetic drift during such periods was widely recognized in
literature \citep[see e.g.][]{Jokipii77,JokipiiKopriva1979,Potgieter85,BoellaEtAl2001,Strauss2011,DellaTorre2012,DellaTorre2013AdvAstro,ICRC13_DellaTorre}.~
\par
The latest low solar activity period was investigated, in particular, using data taken by PAMELA~\citep[e.g., see][]{PamelaProt2013}.~Previously, AMS-01 mission~\citep[June 1998,][]{AMS01_prot} on the space shuttle
and BESS \citep{bess_prot,BESS2007_Abe_2016} on board of stratospheric balloons, sample few short time periods, during solar cycle 23.~In Fig.~\ref{fig:ProtonHelium_Low}, we show the comparison
between experimental data from PAMELA and modulated spectra from \helmod{} for both protons and helium nuclei.~It has to be remarked that, at rigidities lower than 1.5\,GV, the observed helium spectrum is slightly lower than that expected by the modulated one.
\par
Furthermore,
%
in Fig.~\ref{fig:PAMELATime_Low}, we show the time evolution of a few single rigidity bins observed by PAMELA from 2006 to 2010 during the solar minimum~\citep[see also][and reference therein]{2014SoPh..289..391P,2016SoPh..291.2181V}.~The \helmod{} modulated intensity evaluated for the same period and rigidity is superimposed with solid line. As described in Sect.~\ref{Sec::K0Fit},
in \helmod{} the diffusion coefficient scales along the time using a practical relationship with smoothed sunspot numbers.~The good agreement observed in Fig.~\ref{fig:PAMELATime_Low} confirms that, for quiet periods, even if sporadic solar events can perturb the interplanetary particle transport, the latter is mostly regulated by its average properties.
%

\begin{figure*}[htb]
\centerline{
 \includegraphics[width=0.8\textwidth]{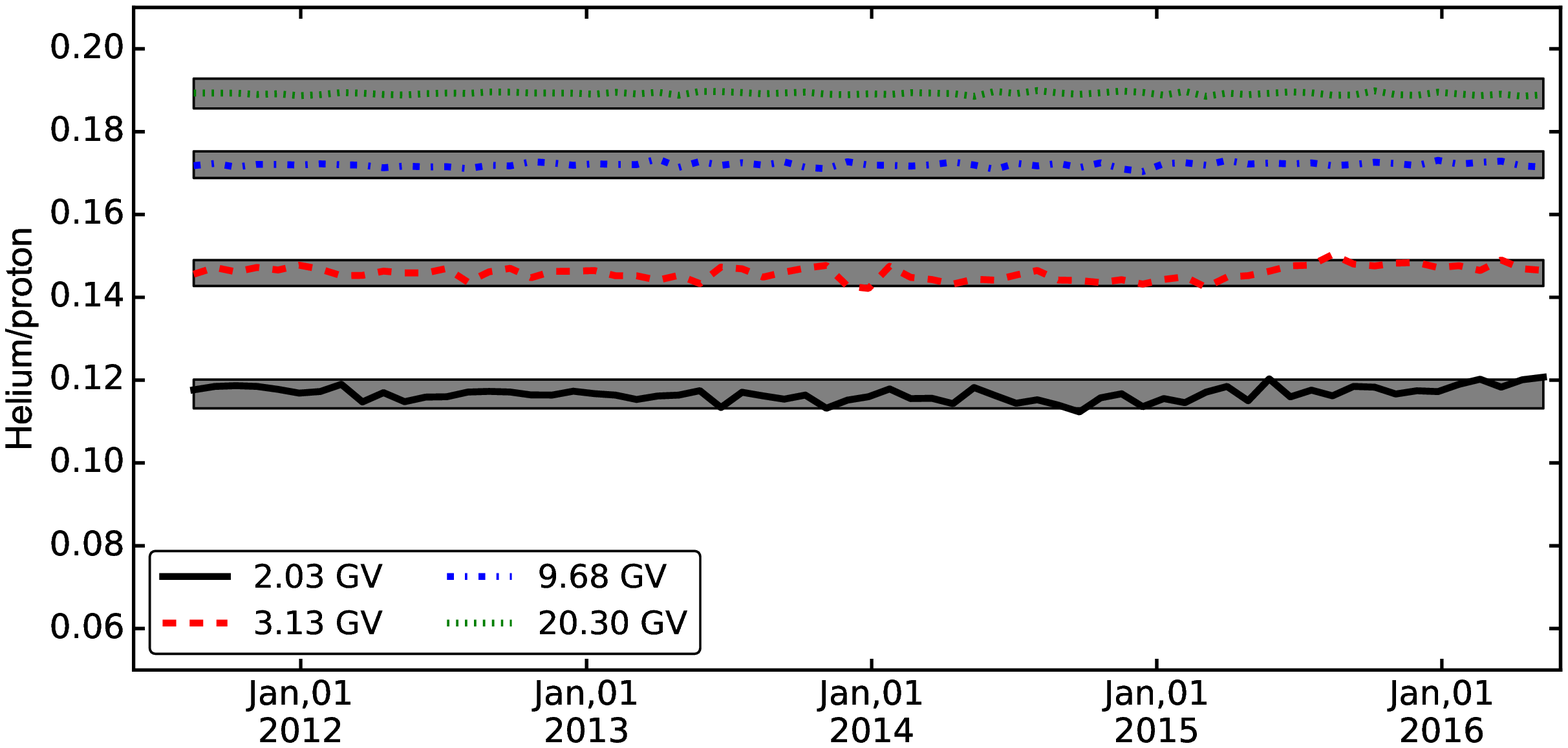}
 }
 \caption{Ratio between Helium and Proton flux in Bartel rotation from June 2011 to May 2016 bins for 2 GV (black solid line), 3 GV (red dashed line), 10 GV (blue dash-dot line) and 20 GV (green dashed line)  as they should be observed by AMS-02 detector. The gray area represents the error of helium over proton ratio as obtained from AMS-02 integral spectrum  (see text).}
 \label{fig:HepRAtion}
\end{figure*}
%
\subsection{High Solar Activity}\label{Sect::HighSolar}
The description of the transport properties during high solar activities is more challenging in comparison with those at low.~The higher rate of solar events makes the interplanetary medium properties more complex and, in turn, the IMF far from being modeled with a regular shape.~
\par
The lack of systematic observations of GCR fluxes in high solar activity makes AMS-02 the first detector, which is able to obtain unprecedentedly precise and continuous measurements of GCRs, under such solar conditions.~Previously, only BESS balloons \citep{bess_prot}
provided proton spectra for short-time periods during the peak of solar cycle 23.~On other hands, AMS-02 already provided an unique data-set integrated over 3 year~\citep{AMS02_2015_PhysRevLett1,AMS02_2015_PhysRevLett2}
of data taking, during the solar activity peak of solar cycle 24.
\par
Recently~\citep[see,][]{ECRS2016a,ECRS2016b,LIS_ApJ2017} \helmod{} model extended his results to reproduce AMS-02 flux of protons, helium nuclei and antiprotons integrated from 2011 to 2015, allowing one to begin unveiling the solar maximum period with an unprecedented detailed treatment.~As
discussed in Sect.~\ref{SEC::kpar}, \helmod{} propagation model for high activity periods needed some additional refinement.~First of all, the higher rate of solar particle emission makes hard to describe the interplanetary medium by means of its average properties only.~In order to overcome this difficulty, NMCR was used instead of SSN as a proxy for scaling diffusion parameter's time variation (see discussion in Sect.~\ref{SEC::kpar}).~Moreover, as discussed in Sect.~\ref{SEC::Parameters},
we introduced an additional correction factor that suppresses drift velocity at the solar maximum.~These improvements allowed \helmod{} to  reproduce the average proton
and helium nuclei fluxes measured by AMS-02 (see Fig.~\ref{fig:pHe_High}) and by BESS~\citep{ECRS2016b,LIS_ApJ2017}, within the experimental error bars and the simulation uncertainties.
\par
So far, it is important to remark that using AMS-02 data we could explore the solar maximum as a whole.~When AMS-02 time dependent data will be available a deeper understanding of the particle transport in solar maximum conditions will be possible.
\subsection{Helium over Proton Ratio}\label{Sect::Hep}
The helium over proton ratio can provide useful insights for understanding the propagation of cosmic rays.~
\par
In Fig.~\ref{fig:HepRAtion}, we made a prediction for helium over proton ratios at different rigidities, as they should be observed by AMS-02 detector from June 2011 to May 2016.~From an inspection of Fig.~\ref{fig:HepRAtion}, one can observe that the ratio is almost constant with time for all considered rigidities.~Helium has double charge and approximately four time the mass of protons, thus processes described in Eq.~\eqref{eq_parker} produce a different level of solar modulation for similar kinetic energy per nucleon.~In fact, diffusion process [see Eq.~\eqref{EQ::KparActual}], magnetic drift [see Eq.~\eqref{drift_antisym_ten}] and adiabatic energy loss (latter term in Eq.~\ref{eq_parker_p}) are naturally expressed in term of particle rigidity ($P=\frac{pc}{Ze}$).~As a consequence, solar modulation acts in the same way for particle with same rigidity and charge sign.~Furthermore, this should occur in spite of the large intensity variations as function of time for the proton fluxes -- in particular at
rigidities lower than about 5--6\,GV --, observed by AMS-02 (e.g., see \citealp{Con2016}).~Thus, particle rigidity is the natural quantity for studying GCR propagation.
\par
Finally, it is important to remark that, in case of particles with opposite charge sign, processes like the magnetic drift transport result in differently propagating positive charged and negative charged particles.~Therefore, a different time behavior is expected to be observed and should be accounted for by accurate propagation models~\citep[e.g., see][]{DellaTorre2012,2016PhRvL.116x1105A}.
\subsection{Dependence on Heliospheric Latitude}\label{Heliosph_latid_dep}
%
\begin{figure*}[hbt]
\centerline{
        \includegraphics[width=0.9\textwidth]{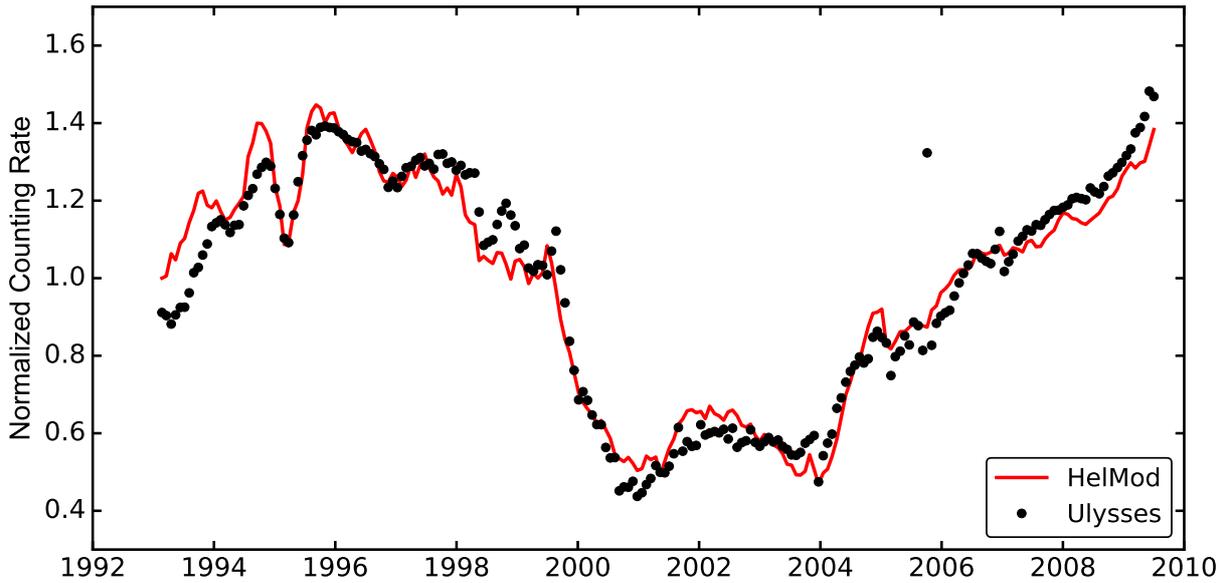}
        }
\caption{Helium normalized counting rate measured by Ulysses (full black
circles) at $\pm 80\degree$ of solar latitude and 1 to 5\,AU compared with the 1\,GeV energy modulated spectrum from \helmod{} code (red solid line) as function of time.}
\label{fig:UlyssesHe}
\end{figure*}
A unique set of measurements of heliosphere
properties out of the ecliptic plane were performed from 1990 to 2009 by the Ulysses mission
(see \citealp{Heber2011} or \citealp{Heber2006} for recent
reviews).~The COSPIN suite of instruments on board of such a spacecraft --
described by \citet{Simpson1992} -- includes the Kiel electron
telescope (KET), designed to measure intensities and energy
spectra of energetic particles separating electrons, protons and
helium nuclei.~The instrument covered the energy range from $\sim$5\,MeV/nucleon to
above 2\,GeV/nucleon for protons and helium nuclei, and from $\sim3$\,MeV to above
300\,MeV for electrons.~Ulysses orbited around Sun reaching a
maximum solar latitude of about $80^{\circ}$ in both the northern and
southern hemispheres, at solar distances ranging from $\sim 1$\,AU to
$\sim 5$\,AU and with an orbital period of approximately 5.5
years.~Ulysses performed three ``fast latitude scans'' (FLS) of
Sun.~The first FLS in 1994/1995 took place near solar minimum with
positive polarity ($A>0$).~A non-symmetric galactic proton
intensity with respect to the heliographic equator was observed.~The
minimal intensity for protons with energies $>0.1$\,GeV was
observed to be displaced by about 7--10 degrees towards the southern
hemisphere \citep{McKibben1996,Simpson1996a,Heber1996a}.~Moreover
latitudinal gradients of ~0.3\%/degree for protons at low energies
($<0.1\,$GeV) and 0.22\%/degree at higher energies ($>2\,$GeV) were
observed.~
\par
The second FLS in
2000/2001 took place close to solar maximum with negative polarity
($\rm A<0$) and no latitudinal gradient was observed for any
GCR species, indicating that drift effects are negligible at solar
maximum~\citep{McKibben2003}.
\par
The third FLS in 2007/2008 was performed close to solar
minimum, as the first FLS, but with reversed polarity ($A<0$).~Observations made by PAMELA starting from 2006
(\citealt{PAMELA})
allowed a comparative determination of the radial and
latitudinal gradient during an $A<0$ solar magnetic epoch.~For protons in the rigidity interval 1.6--1.8\,GV the
measured latitudinal gradient found was ($-0.024 \pm 0.005$)\%/degree and
the radial gradient ($2.7 \pm 0.2$)\%/AU~\citep{deSimone2011}.~These measurements, performed during low solar
activity, allowed a deeper insight into drift mechanisms
and the structure of the heliospheric magnetic field in the polar regions.~In this way, it was possible to test IMF models like the one proposed by \citet{Fisk1996}.~
\par
Detailed studies with \helmod{} model using Ulysses results explored
the solar modulation outside the ecliptic plane.~In \citet{DellaTorre2013AdvAstro}, it was proven that,
using a propagation model as the one described in Sect.~\ref{Model},
\helmod{} code is able to reproduce qualitatively and quantitatively the latitudinal profile of the GCR
intensity, and the latitudinal dip shift with respect to the ecliptic
plane as observed in the inner part of heliosphere by the Ulysses spacecraft
during the first FLS.~As an illustrative example, in Fig.~\ref{fig:UlyssesHe} we show the comparison between the measured helium normalized counting rate in the energy range (0.250-2.1)\,GeV and the modulated spectrum calculated using \helmod{} for 1\,GeV.
Both experimental data and simulations are normalized to the mean value to allow a relative comparison along the solar cycle.~It is important  to remark that, the aim of Fig.~\ref{fig:UlyssesHe} is to show the qualitative agreement found between the \helmod{} spectra and observation data; in fact, \helmod{} calculations were performed for a mono-energetic bin, while KET observations are integrated over a large energy interval~\citep[e.g., see the discussion in][]{deSimone2011}.~A more quantitative comparison with Ulysses data might need to combine together simulations for several energy bins weighted with Ulysses response function.~The
first FLS in 1994/1995 and the third in 2007/2008 are almost fully
reproduced.~Besides, the second FLS in 2000/2001 during the solar maximum still shows
discrepancies, which deserve to be better understood.
\section{Conclusions}
\label{Cncl_sect}
The CR propagation inside the heliosphere was initially treated -- almost 60 years ago -- by~\citet{parker1965} who provided a general theoretical framework for the heliospheric modulation through the interplanetary medium.~Since then, continuous advances allowed a deeper and deeper understanding of the general properties of the IMF affecting particle motion.~However, the description of the transport mechanisms occurring still needs further refinements for allowing a comprehensive treatment along the complete solar cycle.~Moreover, since the heliosphere cannot be considered a steady-state environment, a model should be able to reproduce in a unique description both low and high activity periods.~The idea behind \helmod{} is, in fact, that since all GCR species propagating in the heliosphere experience the same mechanisms, then the modulated spectra can be derived using the same time-dependent heliospheric parameters, i.e., the entire set of modulated observed spectra has to be reproduced with those
parameters from the corresponding LIS's.
\par
In the present work, we have treated the heliospheric parameters of \helmod{} model and how they were implemented in the code for the version 3, as well the Monte Carlo integration technique with a full description of the SDEs, which finally allow the numerical solution of the FPE proposed by~\citet{parker1965}.
\par
A relevant step forward in this field is currently due to the availability of data sets, regarding the time evolution of modulated CR spectra -- provided by PAMELA and AMS-02 spectrometers -- from the final part of the solar cycle 23 up to present.~The numerical approach of \helmod{} model version 3 -- discussed in this work and in the recent article by \citet{ECRS2016a,ECRS2016b,LIS_ApJ2017} -- is able to satisfactorily reproduce the full set of observations obtained, during the two latest solar cycles, for instance by BESS, AMS-01, PAMELA and AMS-02.~It has to be pointed out that the unprecedented accuracy provided by AMS-02 proton data with experimental error down to 1--2\% represents a challenge for a modulation model.~The agreement found between proton and helium nuclei modulated spectra from AMS-02 and those calculated from \helmod{} is within the AMS-02 errors.
\par
Although diffusion is the dominant modulation process for GCRs, it was shown how charge sign dependent processes, i.e., those resulting in a drift velocity convection, are fundamental elements for dealing with a complete solution for the particle transport through the heliosphere.~Moreover, it was remarked that when the treatment accounts only for spectra observed at Earth, it can lead not to appropriately describe the complexity of the modulation phenomenon.~Probes, like Ulysses, allow one to better investigate how to deal with a bi-dimensionally modeled heliosphere, i.e., the modulation effects occurring also out of the ecliptic plane as a function of solar latitude.~For instance, the measured helium normalized counting rate from Ulysses and the modulated spectrum calculated using \helmod{} were found to agree, thus indicating the capabilities of the present model to investigate the CR transport up to large solar latitudes and distances from Sun up to 5\,AU.
\par
Finally, a proper description of heliospheric modulation, like that \helmod{} can provide, relates not only computed LIS's to observed modulated differential intensities, but also may result in constraining the parameters of galactic production models, like \galprop{}~\citep[as discussed in][]{ECRS2016b,LIS_ApJ2017}.
\section*{Acknowledgements}

This work is supported by ASI (Agenzia Spaziale Italiana) under contract ASI-INFN I/002/13/0 and ESA (European Space Agency) contract 4000116146/16/NL/HK.
\par
We acknowledge the NMDB database (www.nmdb.eu), supported under the European Union's FP7 programme (contract no. 213007) for providing data.~The data from McMurdo were provided by the University of Delaware with support from the U.S. National Science Foundation under grant ANT-0739620.~Finally, we acknowledge the use of NASA/GSFC's Space Physics Data Facility's OMNIWeb service, and OMNI data.
\par
We wish to specially thank Pavol Bobik, Giuliano Boella, Davide Grandi, Karel Kudela, Simonetta Pensotti, Marian Putis, Davide Rozza, Mauro Tacconi and Mario Zannoni for their support to \helmod{} code and suggestions.
\appendix
%
%
\section{Diffusion Tensor and Stochastic Differential Equations}
\label{app1}
One of the key point to provide a proper stochastic integration is to evaluate
the correct set of SDEs.~Although the procedure is relatively simple and straightforward,
there are few points that are important to be remarked.~In this section, we evaluate SDEs from
Parker's equation in the form of particle density for unit space and energy and omni-directional distribution function.~More details on the procedure can be found in~\citet{Zhang1999,PeiBurger2010,Kopp2012,DellaTorre2016_OneD} and reference therein.
\subsection{SDE in Kinetic Energy}\label{App:SDETKin}
Equation~\eqref{eq_parker}, that controls cosmic rays modulation, can be
always written by a multi-dimensional diffusion equation in the form of
backward-in-time Kolmogorov equation~\citep{Zhang1999}:
\begin{equation}\label{eq::KolmBackGeneral}
\frac{\partial F}{\partial t} =\frac{1}{2}\sum_{i,j}[ B  B^\top]^{ij} \frac{\partial^2 F }{\partial x_i \partial x_j}
                                + \sum_{i} A_B^i \frac{\partial F}{ \partial x_i}
                                + L F.
\end{equation}
The corresponding SDE set is, then, obtained as:
\begin{equation}
 dx_i = A_B^i dt + \sum_j B^{ij} d\omega_j.
\end{equation}
It is important to note that Eq.~\eqref{eq::KolmBackGeneral} includes an additional parameter $L$,
that is not directly taken into account for the evaluation of the stochastic path.~Indeed, it represents an additional process by allowing the stochastic realization to be created at an exponential rate of $L$ as function of time (see Eq.~\ref{eq::JbackT}).~Equation~\eqref{eq_parker}, in a spherical heliocentric coordinate system, can be rewritten as:
\begin{align}\label{eq::kolmoxSDE}
 \frac{\partial U}{\partial t}=& \frac{1}{r^2}\frac{\partial r^2K_{rr}}{\partial r} \frac{\partial U}{\partial r}
				 +\frac{1}{r\sin\theta}\frac{\partial}{\partial \theta}\left(\sin\theta K_{\theta r}\right)\frac{\partial U}{\partial r}
\nonumber\\		       & + \frac{1}{r\sin\theta}\frac{\partial K_{\phi r}}{\partial \phi} \frac{\partial U}{\partial r}
				 - V_{sw} \frac{\partial U}{\partial r}
				 - V_{dr} \frac{\partial U}{\partial r}
\nonumber\\
			      & + \frac{1}{r^2}\frac{\partial rK_{r\theta}}{\partial r} \frac{\partial U}{\partial \theta}
				 + \frac{1}{r\sin\theta}\frac{\partial}{\partial\theta}\left(\frac{K_{\theta \theta}\sin\theta}{r}\right)  \frac{\partial U}{\partial \theta}
\nonumber\\		       & + \frac{1}{r^2\sin\theta}\frac{\partial}{\partial\phi}\left(K_{\phi \theta}\right) \frac{\partial U}{\partial \theta}
				 - \frac{V_{d\theta}}{r} \frac{\partial U}{\partial \theta}
\nonumber\\
			      & + \frac{1}{r^2}\frac{\partial}{\partial r}\left(\frac{ rK_{r\phi}}{\sin\theta}\right) \frac{\partial U}{\partial \phi}
				 + \frac{1}{r\sin\theta}\frac{\partial}{\partial \theta} \left( \frac{K_{\theta\phi}}{r}\right) \frac{\partial U}{\partial \phi}
\nonumber\\		       & + \frac{1}{r^2\sin^2\theta}\frac{\partial}{\partial \phi }\left(K_{\phi \phi}\right)\frac{\partial U}{\partial \phi}
				 - \frac{V_{d\phi}}{r\sin\theta} \frac{\partial U}{\partial \phi}
				 + \frac{2}{3} \frac{\alpha T V_{sw}}{r} \frac{\partial U}{\partial T}
\nonumber\\
			      & + K_{rr}\frac{\partial^2 U}{\partial r \partial r}
				 + \frac{K_{r\theta}}{r}\frac{\partial^2 U}{\partial r \partial \theta}
				 + \frac{K_{r\phi}}{r\sin\theta} \frac{\partial^2 U}{\partial r \partial \phi}
\nonumber\\
                              & + \frac{K_{\theta r}}{r}\frac{\partial^2 U}{\partial \theta \partial r}
                                 + \frac{K_{\theta \theta}}{r^2}\frac{\partial^2 U}{\partial \theta^2}
				 +\frac{K_{\theta\phi}}{r^2\sin\theta}\frac{\partial^2 U}{\partial \theta \partial \phi}
\nonumber\\
                              & + \frac{K_{\phi r}}{r \sin \theta}\frac{\partial^2 U}{\partial \phi \partial r}
                                 + \frac{K_{\phi \theta}}{r^2 \sin \theta} \frac{\partial^2 U}{\partial \phi \partial \theta}
                                 + \frac{K_{\phi \phi}}{r^2 \sin^2\theta} \frac{\partial^2 U}{\partial \phi^2 }
\nonumber\\
			      & - \frac{1}{r^2}\frac{\partial r^2 V_{sw}}{\partial r} U
                                + \frac{2}{3} \frac{V_{sw}}{r}\frac{\partial \alpha T}{\partial T} U,
\end{align}
where the divergence of drift velocity is zero by definition:
\[  - \frac{1}{r^2}\frac{\partial r^2 V_{dr}}{\partial r}
                                 - \frac{1}{r\sin\theta}\frac{\partial\sin\theta V_{d\theta}}{\partial \theta}
                                 - \frac{1}{r\sin\theta}\frac{\partial V_{d\phi}}{\partial \phi}  = \nabla \cdot V_d = 0 . \]
From a comparison between Eq.~\eqref{eq::KolmBackGeneral} and Eq.~\eqref{eq::kolmoxSDE}, it is possible to obtain:
\begin{eqnarray}\label{eq::SDE_backward}
A^r_B &=& \frac{1}{r^2}\frac{\partial r^2K_{rr}}{\partial r} +\frac{1}{r\sin\theta}\frac{\partial}{\partial \theta}\left(\sin\theta K_{\theta r}\right) \nonumber \\
      &&  + \frac{1}{r\sin\theta}\frac{\partial K_{\phi r}}{\partial \phi} - V_{sw} - V_{dr} \\
A^\theta_B &=&\frac{1}{r^2}\frac{\partial rK_{r\theta}}{\partial r}  + \frac{1}{r\sin\theta}\frac{\partial}{\partial\theta}\left(\frac{K_{\theta \theta}\sin\theta}{r}\right) \nonumber \\
      &&+ \frac{1}{r^2\sin\theta}\frac{\partial}{\partial\phi}\left(K_{\phi \theta}\right)- \frac{V_{d\theta}}{r}  \\
A^\phi_B &=&\frac{1}{r^2}\frac{\partial}{\partial r}\left(\frac{ rK_{r\phi}}{\sin\theta}\right) + \frac{1}{r\sin\theta}\frac{\partial}{\partial \theta} \left( \frac{K_{\theta\phi}}{r}\right) \nonumber \\
      &&+\frac{1}{r^2\sin^2\theta}\frac{\partial}{\partial \phi }\left(K_{\phi \phi}\right) - \frac{V_{d\phi}}{r\sin\theta}\\
A^T_B &=&\frac{2}{3} \frac{\alpha T V_{sw}}{r}\\
 B  B^\top&=& \left[\begin{array}{ccc}
                           2 K_{rr} 				&\frac{2K_{r\theta}}{r}		&\frac{2K_{r\phi}}{r\sin\theta} \\
			    \frac{2K_{\theta r}}{r}	&\frac{2K_{\theta \theta}}{r^2}	&\frac{2K_{\theta\phi}}{r^2\sin\theta}\\
                             \frac{2K_{\phi r}}{r \sin \theta}	&\frac{2K_{\phi \theta}}{r^2 \sin \theta}&\frac{2K_{\phi \phi}}{r^2 \sin^2\theta}	
                           \end{array} \right]\\
L&=&   \frac{2V_{sw}}{r} \left (\frac{1}{3}\frac{\partial \alpha T}{\partial T} -1 \right).
\end{eqnarray}
The matrix $B$ can be derived using the Cholesky--Banachiewicz or Cholesky--Crout algorithms.~As pointed out in Appendix B of~\cite{PeiBurger2010}, the so obtained matrix is not unique,
but its solutions are stochastically equivalent~\citep[see e.g.][]{Kopp2012}.
\subsection{FPE in momentum}
To obtain SDE from Eq.~\eqref{eq_parker_p}, the latter should be rewritten in the form of Eq.~\eqref{eq::KolmBackGeneral}
as described in~\ref{App:SDETKin}:
{\small
\begin{eqnarray}\label{eq::SDE_backward_f}
A^r_B &=& \frac{1}{r^2}\frac{\partial r^2K_{rr}}{\partial r} +\frac{1}{r\sin\theta}\frac{\partial}{\partial \theta}\left(\sin\theta K_{\theta r}\right) \nonumber \\
      &&  + \frac{1}{r\sin\theta}\frac{\partial K_{\phi r}}{\partial \phi} - V_{sw} - V_{dr} \\
A^\theta_B &=&\frac{1}{r^2}\frac{\partial rK_{r\theta}}{\partial r}  + \frac{1}{r\sin\theta}\frac{\partial}{\partial\theta}\left(\frac{K_{\theta \theta}\sin\theta}{r}\right) \nonumber \\
      &&+ \frac{1}{r^2\sin\theta}\frac{\partial}{\partial\phi}\left(K_{\phi \theta}\right)- \frac{V_{d\theta}}{r}  \\
A^\phi_B &=&\frac{1}{r^2}\frac{\partial}{\partial r}\left(\frac{ rK_{r\phi}}{\sin\theta}\right) + \frac{1}{r\sin\theta}\frac{\partial}{\partial \theta} \left( \frac{K_{\theta\phi}}{r}\right) \nonumber \\
      &&+\frac{1}{r^2\sin^2\theta}\frac{\partial}{\partial \phi }\left(K_{\phi \phi}\right) - \frac{V_{d\phi}}{r\sin\theta}\\
A^p_B &=&\frac{2}{3} \frac{p V_{sw}}{r}\\
 B  B^\top&=& \left[\begin{array}{ccc}
                           2 K_{rr} 				&\frac{2K_{r\theta}}{r}		&\frac{2K_{r\phi}}{r\sin\theta} \\
			    \frac{2K_{\theta r}}{r}	&\frac{2K_{\theta \theta}}{r^2}	&\frac{2K_{\theta\phi}}{r^2\sin\theta}\\
                             \frac{2K_{\phi r}}{r \sin \theta}	&\frac{2K_{\phi \theta}}{r^2 \sin \theta}&\frac{2K_{\phi \phi}}{r^2 \sin^2\theta}	
                           \end{array} \right]\\
L&=&0.
\end{eqnarray}
}
One should note that in this case a) the linear term $L$ is equal to zero, thus the exponential weight in Eq.~\eqref{eq::JbackT}
can be neglected~\citep[as done, e.g., in][]{Strauss2011}, b) both spatial convection and diffusion matrix are the same as for SDE in kinetic energy, and c) the difference between SDE in kinetic energy and momentum is only in the energy loss terms: $A^T_B$, $A^p_B$ and $L$.


\section*{References}


\clearpage


\end{document}